\documentclass[aps,nofootinbib,showpacs]{revtex4}
\usepackage{hyperref}
\usepackage{bm}
\usepackage{graphicx}
\usepackage{amsmath}
\usepackage{epsf}
\usepackage{amssymb}
\usepackage{txfonts}
\usepackage{color}

\def\be{\begin{equation}}
\def\ee{\end{equation}}
\def\bea{\begin{eqnarray}}
\def\eea{\end{eqnarray}}
\def\ellt{\ell_1\ell_2\ell_3}
\newcommand{\kv}{{\bf k}}
\newcommand{\tQ}{\tilde Q}
\newcommand{\tU}{\tilde U}

\newcommand{\calb}{a}
\newcommand{\rot}{\omega}

\newcommand{\cmb}{\Theta}

\newcommand{\intlnp}{\int \frac{d^2 {\bf l'}}{(2\pi)^2}}
\newcommand{\intlnpp}{\int \frac{d^2 {\bf l''}}{(2\pi)^2}}

\newcommand{\intlp}{\int {d^2 {\bf l'}\over (2\pi)^2}}
\newcommand{\intlpp}{\int {d^2 {\bf l''}\over (2\pi)^2}}

\newcommand{\bfl}{{\mathbf{l}}}

\newcommand{\bflp}{{\mathbf{l^{\prime}}}}
\newcommand{\bflpp}{{\mathbf{l^{\prime\prime}}}}

\newcommand{\bfL}{{\mathbf{L}}}

\def\fnll{f_{\rm NL}^{\rm loc.}}

\def\fnle{f_{\rm NL}^{\rm eq.}}

\newcommand{\bfx}{{\mathbf{x}}}

\newcommand{\bfk}{{\mathbf{k}}}

\newlength{\tskip}
\newlength{\colwidth}

\newcommand{\beq}{\begin{equation}}
\newcommand{\eeq}{\end{equation}}
\newcommand{\beqa}{\begin{eqnarray}}
\newcommand{\eeqa}{\end{eqnarray}}

\newcommand{\bi}{B_{l_1 l_2 l_3}}

\newcommand{\bn}{\hat{\bf n}}
\newcommand{\bl}{\hat{\bf l}}

\newcommand{\vl}{{\mathbf{l}}}
\newcommand{\nn}{\nonumber\\}
\newcommand{\wthrj}[6]{\left(
                           \begin{array}{ccc}
        \! #1\! & #2\!  & #3\!  \\
        \! #4\! & #5\!  & #6\!
                           \end{array}
               \right)}
\newcommand{\wsixj}[6]{\left\{
                           \begin{array}{ccc}
         #1 & #2  & #3  \\
         #4 & #5  & #6
                           \end{array}
               \right\}}
\newcommand{\mpr}{m^{\prime}}
\newcommand{\mdp}{m^{\prime \prime}}
\newcommand{\mtp}{m^{\prime \prime \prime}}

\newcommand{\mtwopr}{m_2^{\prime}}
\newcommand{\mtwodp}{m_2^{\prime \prime}}

\newcommand{\mthrpr}{m_3^{\prime}}
\newcommand{\mthrdp}{m_3^{\prime \prime}}
\newcommand{\mthrtp}{m_3^{\prime \prime \prime}}
\newcommand{\lp}{l^{\prime}}
\newcommand{\ldp}{l^{\prime \prime}}
\newcommand{\ltp}{l^{\prime \prime \prime}}

\newcommand{\ltwop}{l_2^{\prime}}
\newcommand{\ltwodp}{l_2^{\prime \prime}}

\newcommand{\lthrp}{l_3^{\prime}}
\newcommand{\lthrdp}{l_3^{\prime \prime}}
\newcommand{\lthrtp}{l_3^{\prime \prime \prime}}
\newcommand{\fnl}{{f_{\rm NL}}}

\begin{document}

\title {Impact of Instrumental Systematics on the CMB Bispectrum}

\author{Meng Su$^1$}\email{mengsu@cfa.harvard.edu}
\author{Amit P.S. Yadav$^{1,2}$}
\author{Meir Shimon$^3$}
\author{Brian G. Keating$^3$}

\affiliation{$^1$Institute for Theory and Computation, Harvard-Smithsonian Center for Astrophysics, 60 Garden Street, MS-10, Cambridge, MA 02138 USA}
\affiliation{$2$ Institute for Advanced Study, Princeton, NJ 08540, USA}
\affiliation{$3$ Center for Astrophysics and Space Sciences, University of California, San Diego, La Jolla, CA, 92093, USA}



\begin{abstract}
We study the effects of instrumental systematics on the estimation of primordial non-Gaussianity using the cosmic microwave background (CMB) bispectrum from both the temperature and the polarization anisotropies. For temperature systematics we consider gain fluctuation and beam distortions. For polarization we consider effects related to known instrumental systematics: calibration, pixel rotation, differential gain, pointing, and ellipticity of the intrument beam.  We consider these effects at the next to leading order, which we refer to as non-linear systematic effects.  We find that if the instrumental response is linearly proportional to the received CMB intensity, then only the shape of the primordial CMB bispectrum, if there is any, will be distorted. We show that the nonlinear response of the instrument can in general result in spurious non-Gaussian features on both the CMB temperature and polarization anisotropies, even if the primordial CMB is completely Gaussian.  We determine the level for both the linear and non-linear systematics parameters for which they would cause no significant degradation of our ability to constrain the primordial non-Gaussianity amplitude $\fnl$. We find that the non-linear systematics are potentially bigger worry for extracting the primordial non-Gaussianity than the linear systematics. Especially because the current and near future CMB probes are optimized for CMB power-spectrum measurements which are not particularly sensitive to the non-linear instrument response. We find that if instrumental non-linearities are not controlled by dedicated calibration, the effective local non-Gaussianity can be as large as $\fnl\sim O(10)$ before the corresponding non-linearities show up in the CMB dipole measurements. The higher order multipoles are even less sensitive to instrumental non-linearities.   
\end{abstract}

\maketitle

\section{Introduction}
\label{sec:intro}

Characterizing the non-Gaussianity in the primordial perturbations has emerged as a powerful probe of the early universe. Different inflationary models predict different non-Gaussian signal in both amplitude and shape (see review~\cite{2009astro2010S.158K}). The specific departures from Gaussianity are highly model-dependent and any detection would rule out the simplest single field inflationary models. Non-Gaussianities measures the strength of interaction during inflation and can be studied and characterized most directly by using the effective theory approach ~\cite{Cheung2008,Weinberg2008,SZ10}. The first statistic that 
captures non-Gaussianity
is the three-point correlation function or bispectrum in Fourier space. The overall amplitude of non-Gaussianity constrained from the data is often quoted in terms of a dimensionless non-linearity parameter $\fnl$, and a shape which specify the configurations of the wavevectors of perturbations that contain the highest contributions to the non-Gaussian signal. With the assumptions of translational, and rotational invariance, the bispectrum shapes can be characterized as different triangle shapes. The three most studied shapes are: 
the `local shape' where the non-Gaussian signal is maximum on squeezed configurations ($k_1 \ll k_2,k_3$), the `equilateral shape' where the bispectrum peaks mostly on equilateral triangles ($k_1 \sim k_2 \sim k_3$), and the `orthogonal shape' which peaks for both equilateral and flat-triangle configurations (see e.g.~\cite{Chen2010,Liguori2010} for a review).

Different models of inflation predict different levels of $\fnl$, 
ranging from $O(1)$ to $\fnl\sim O(100)$. Values above $\fnl\sim O(100)$ have been 
ruled out
by the WMAP data already. Non-Gaussianity from 
``classical" inflation models that are based on a slowly rolling scalar field is very small~\cite{Acquaviva02,Maldacena03}; however, a very large class of more general models with, e.g., multiple scalar fields, features in inflaton potential, non-adiabatic fluctuations, non-canonical kinetic terms, deviations from Bunch-Davies vacuum, among others generates 
significantly
higher 
levels
of non-Gaussianity. Equilateral shapes are generically a signature of nonstandard kinetic terms in the inflaton Lagrangian, as for example in DBI \cite{Alishahiha2004} and ghost inflation \cite{Arkani_et_04}. 

Detections of any type of primordial non-Gaussianity would have profound implications on our understanding of the early Universe
and it is therefore crucial to estimate all
possible contaminations of non-Gaussianity in CMB. Any physical/instrumental systematics to the non-Gaussian measurement must be well understood and controlled. We have to 
ascertain
that any detected non-Gaussian signal has a {\em physical origin} and is not 
spurious.
Potential contaminants include residual foreground contamination and unresolved point sources, non-Gaussianity induced by second-order anisotropies, such as gravitational lensing, the Sunyaev-Zel'dovich effect, and 
non-uniform
recombination \cite{Yadav2007,Smith2009}. 

In this paper, a detailed study of the impact of instrumental systematics on the primordial bispectrum measurement from CMB temperature and polarization fields is provided. We first show how such systematics change the CMB bispectrum signal for both temperature and polarization fields. We show that apart from distorting the primordial bispectrum, non-linearities in the instrument can also generate spurious bispectrum even in the absence of primordial bispectrum. This shape 
distortion
could modify the effective normalization of a $\fnl$ estimator 
or even confuse the various primordial shapes of the bispectra.
For a reliable inference of the level of primordial non-Gaussianity from upcoming CMB data, and thereby probe of the early Universe physics, precise and realistic estimation of instrumental systematics are required.

The remainder of this paper is organized as follows. In \S\ref{sec:cmbbispectrum} we introduce the basics of CMB bispectrum calculation. In \S\ref{sec:systematics} we present the parametrization of different instrumental systematics and explain the relations to the real experimental effects. In \S\ref{sec:distortions} we detail the modeling of linear and non-linear calibration systematics of CMB temperature anisotropies, and show that the systematics would generate new bispectrum and distort existing bispectrum. In \S\ref{sec:polabispectrum} we calculate the systematics effects on bispectrum involving CMB $E$-mode polarization. Our numerical results and discussion are given in \S\ref{sec:results}, where we also calculate the bias due to the non-linear calibration, the induced statistical error in $\fnl$ parameters due to non-Gaussian statistics of the distorted CMB fields. We compare the systematics requirement from bispectrum measurement with the requirement derived from the CMB $B$-mode polarization detection in \S\ref{sec:Bmode}. We also discuss our findings in the context of ongoing PLANCK and other upcoming CMB experiment. We summarize and draw our conclusions in \S\ref{sec:summary}. All our instrumental systematics calculations are carried out within the Jones matrix formalism where in Appendix A we describe the alternative, Stokes-parameters-based parametrization of beam distortions. In appendix B we describe CMB bispectra obtained in the full-sky formalism to supplement our calculations which have been carried out under the approximation of flat sky.

\section{The Primordial Non-Gaussianity in the CMB Bispectrum}
\label{sec:cmbbispectrum}

To characterize the non-Gaussianity one has to consider the higher order moments beyond the two-point function, which contains all the information for Gaussian perturbations. The 3-point function, which identically vanishes for Gaussian perturbations, contains information about non-Gaussianity~\cite[for details]{BKMR_04, 2009astro2010S.158K,2010AdAst2010E..75B,2010CQGra..27l4010K,Chen2010,Liguori2010,2010AdAst2010E..71Y}. 
Non-Gaussianities from the early universe can be described by the 3-point correlation function of Bardeen's curvature perturbations, $\Phi(k)$, which can be simplified assuming translational symmetry to
\begin{eqnarray}
\langle \Phi(\mathbf{k_1})\Phi(\mathbf{k_2})\Phi(\mathbf{k_3})\rangle = (2\pi)^3\delta^3(\mathbf{k_1} + \mathbf{k_2} + \mathbf{k_3}) \fnl \cdot F(k_1, k_2, k_3)
\label{eq:3pt}
\end{eqnarray}
where $F(k_1, k_2, k_3)$ describes the shape of the bispectrum in Fourier space while the amplitude of non-Gaussianity is captured by the dimensionless non-linearity parameter
$\fnl$. The shape function $F(k_1,k_2,k_3)$ correlates 
perturbations
with three wave-vectors and form a triangle in Fourier space. Depending on the physical mechanism responsible for the bispectrum, the shape of the 3-point function, $F(k_1, k_2, k_3)$ can be broadly classified into three
classes~\citep{Babich_etal_04,Holman:2007na}.
The local, ``squeezed,'' non-Gaussianity where
$F(k_1, k_2, k_3)$ is large for 
configurations 
where $k_1 \ll k_2\approx k_3$. Next, the ``equilateral,'' non-Gaussianity where $F(k_1, k_2, k_3)$ is
large for configurations where $k_1 \approx k_2 \approx k_3$. Recently, a new bispectrum template shape, an orthogonal shape, has been introduced~\cite{Senatore:2009gt} which characterizes the size of the signal ($f^{ortho}_{\rm NL}$) and peaks at both equilateral and flat-triangle configurations.

It is easy to show that the shape dependence for the local model, $F_{\rm loc.}(k_1,k_2,k_3)$, takes the following form
\be
\label{eq:f_local}
F_{\rm loc.}(k_1, k_2, k_3) = 2 \Delta^2_{\Phi}\left[ \frac{1}{k^{3-(n_s-1)}_1 k^{3-(n_s-1)}_2} + \frac{1}{k^{3-(n_s-1)}_1k^{3-(n_s-1)}_3}+\frac{1}{k^{3-(n_s-1)}_2 k^{3-(n_s-1)}_3}\right],
\ee  
where the normalized power spectrum $\Delta_\Phi$ is defined in terms of the tilt $n_s$ and the curvature power spectrum 
as $P_\Phi(k)\equiv \Delta_\Phi k^{-3+(n_s-1)}$ with $P_\Phi(k)$ defined in terms of the Gaussian component alone, $\langle\Phi_{G}({\bf k}_1) \Phi_{G}({\bf k}_2) \rangle \equiv \delta_D^{(3)}(\kv_{12}) P_\Phi(k_1)$. 

Equilateral forms of non-Gaussianity arise from models with non-canonical kinetic terms such as the DBI action~\citep{Alishahiha_etal04}, ghost condensation~\citep{Arkani_et_04}, or any other single-field models in which the scalar field acquires a low speed of sound~\citep{Chen_etal07,Cheung_Creminelli_etal07}. Although the shapes predicted by different models are in this case not identical, it has been noted~\cite{Babich_etal_04,creminelli_wmap1} that they are all very well-approximated by the function 
\begin{eqnarray}
\label{eq:f_eq}
F_{\rm eq.}(k_1, k_2, k_3) & = &  
6\Delta^2_{\Phi}\left[- \frac{1}{k^{3-(n_s-1)}_1k^{3-(n_s-1)}_2} + {\rm 2~perm.}
       - \frac{2}{(k_1k_2k_3)^{2-2(n_s-1)/3}}\right.
\nonumber\\
& &    \left. + \frac{1}{k^{1-(n_s-1)/3}_1k^{2-2(n_s-1)/3}_2k^{3-(n_s-1)}_3} + {\rm 5~perm.}\right].
\end{eqnarray}
The definition for the equilateral model follows from the local one since $\fnle$ is defined in such a way that for equilateral configurations, $F_{\rm eq.}(k,k,k)=F_{\rm loc.}(k,k,k)$ and one obtains the same value for $B_\Phi$ given $\fnle=\fnll$. The local and equilateral forms are nearly orthogonal to each other, which implies that both can be measured nearly independently.
The orthogonal form of non-Gaussianity is 
nearly orthogonal to both the local and equilateral forms \cite{Senatore:2009gt}
\begin{eqnarray}
\label{eq:f_orth}
F_{\rm ortho}(k_1,k_2,k_3) & = & 6\Delta^2_{\Phi}\left\{-\frac3{k^{4-n_s}_1k^{4-n_s}_2}-\frac3{k^{4-n_s}_2k^{4-n_s}_3}-\frac3{k^{4-n_s}_3k^{4-n_s}_1}\right.
\nonumber\\
& & \left. -\frac8{(k_1k_2k_3)^{2(4-n_s)/3}}+\left[\frac3{k^{(4-n_s)/3}_1k^{2(4-n_s)/3}_2k^{4-n_s}_3} +\mbox{(5 perm.)}\right]\right\}.
\label{eq:Forthog}
\end{eqnarray}
The motivation of this shape is that a certain inflationary models yield a distinct shape which is orthogonal to both the local and equilateral forms. The orthogonal form has a positive peak at the equilateral configuration, and a negative valley along the elongated configurations \cite{Komatsu:2010}. 

All single-field inflation models predict small amplitude for bispectrum in the limit of squeezed configurations, $f^{local}_{\rm NL}=\frac{5}{12}(1-n_s)$, regardless of the form of potential or kinetic term. Given that $1-n_{s}\approx 0.04$, all single field models predict $f^{local}_{\rm NL} \approx 0.02$. Hence the detection of local-type non-Gaussianity with amplitude greater than $O(0.1)$ would essentially rule out {\it all} single-field inflation models. For a credible non-Gaussianity detection, it would be essential to control all effects which might potentially contaminate the bispectrum.

{\it Primordial non-Gaussianity in the CMB:} The harmonic coefficients of the CMB anisotropy $a^X_{lm}$ for temperature or $E$-mode polarization can be related to the primordial Bardeen curvature fluctuation $\Phi(\bfk)$ via
\begin{eqnarray} 
\label{phi_alm}
a^X_{\ell    m}=4\pi (-i)^\ell\int  \frac{d^3k}{(2 \pi)^3} \Phi(\mathbf{k})   \,    g^X_{\ell}(k)   Y^*_{\ell m}(\hat \kv),
\end{eqnarray} 
where $g^X_{\ell}(r)$ is the radiation
transfer function of temperature ($X=T$) or polarization ($X=E$). The most promising way of measuring non-Gaussianities is to study the CMB angular bispectrum, defined as follows 
\be
B^{XYZ}_{\ell_1 \ell_2 \ell_3, m_1m_2m_3}\equiv \langle a^X_{\ell_1 m_1}a^Y_{\ell_2 m_2}a^Z_{\ell_3 m_3} \rangle\,.
\ee   
Using Eq.(\ref{phi_alm}), the angular-averaged bispectrum can be written as
 \begin{eqnarray}
 \label{eq:cmb_bi}
 \bi^{pqr} & = & (4\pi)^3 (-i)^{\ell_1+\ell_2+\ell_3} \sum_{m_1m_2m_3} \left(\begin{array}{ccc} \ell_1 & \ell_2 & \ell_3 \\ m_1 & m_2 & m_3 \end{array} \right)  \int \frac{d^3k_1}{(2 \pi)^3}\frac{d^3k_2}{(2 \pi)^3}\frac{d^3k_3}{(2 \pi)^3} \; Y^*_{\ell_1 m_1}(\hat{\kv}_1) Y^*_{\ell_2 m_2}(\hat{\kv}_2)Y^*_{\ell_3 m_3}(\hat{\kv}_3) 
 \nonumber \\ & & 
 \times g^p_{\ell_1}(k_1)
 g^q_{\ell_2}(k_2) g^r_{\ell_3}(k_3) \; \langle\Phi(\mathbf{k}_1)\Phi(\mathbf{k}_2)\Phi(\mathbf{k}_3) \rangle \;
 \end{eqnarray}
where $\langle\Phi(\mathbf{k}_1)\Phi(\mathbf{k}_2)\Phi(\mathbf{k}_3) \rangle$ is the primordial curvature three-point function as defined in Eq.~(\ref{eq:3pt}) and
the matrix is the Wigner 3J symbol imposing selection rules which makes bispectrum zero unless \\
 (i) $\ell_1 +\ell_2 + \ell_3 =$integer \\
(ii) $m_1 + m_2 + m_3=0$ \\
(iii) $\vert \ell_i -\ell_j \vert \le \ell_k \le \ell_i + \ell_j$ for $i,j,k=1,2,3$.\\
For simplicity, it is customary to define the reduced bispectrum $b^{XYZ}_{\ell_1 \ell_2 \ell_3}$
\begin{eqnarray}
\bi^{XYZ}&=&\sum_{m_1m_2m_3}  \left(\begin{array}{ccc} \ell_1 & \ell_2 & \ell_3 
\\ m_1 & m_2 & m_3 \end{array}\right)   B^{XYZ}_{\ell_1 \ell_2 \ell_3, m_1m_2m_3} \\
&=&\sqrt{\frac{(2 \ell_1+1) (2 \ell_2+1)(2 \ell_3+1)}{4 \pi}} \left(\begin{array}{ccc} \ell_1 & \ell_2 & \ell_3 
\\ 0 & 0 & 0 \end{array}\right)  b^{XYZ}_{\ell_1 \ell_2 \ell_3}.
\end{eqnarray}
Following convention, we define the radial functions $\alpha(r), \beta(r), \delta(r),$ and $\gamma(r)$ for a given co-moving distance $r$ for the temperature and $E$-mode polarization as
\be
\begin{array}{ll}
\alpha^X_\ell(r) \equiv  \frac2\pi \int \!\!\! dk \; k^2 \, g^X_\ell(k) ~j_\ell(k r)\,, &\beta^X_\ell(r) \equiv  \frac2\pi \int \!\!\! dk \; k^{-1} \, g^X_\ell(k) ~j_\ell(k r)\Delta_\Phi\, k^{n_s-1}\,, 
\\ 
\gamma^X_\ell(r) \equiv  \frac2\pi \int \!\!\! dk \; k \, g^X_\ell(k) ~j_\ell(k r) \Delta_\Phi^{1/3}\,k^{(n_s-1)/3}\,, &
\delta^X_\ell(r)  \equiv  \frac2\pi \int \!\!\! dk  \, g^X_\ell(k) ~j_\ell(k r) \Delta_\Phi^{2/3}\,k^{2(n_s-1)/3}.
\end{array}
\ee
In the expression $j_\ell(k r)$ is the Bessel function of order $\ell$. The transfer functions $g^T_{\ell}(k)$ and $g^E_{\ell}(k)$ are numerically calculated using publicly available codes such as CMBfast~\cite{cmbfast} and CAMB~\cite{camb}.

The bispectrum for the local case can be simply written as
\be
 b^{XYZ}_{\ell_1 \ell_2 \ell_3} =  2 \int_0^\infty \!\!r^2 dr \left[ -\alpha^X_{\ell_1}(r) \beta^Y_{\ell_2}(r) \beta^Z_{\ell_3}(r) 
+ 2 \;{\rm perm.}\right]\,. 
\ee
For equilateral template,
\bea
 b^{XYZ}_{\ell_1 \ell_2 \ell_3} = 6 \int\!\!r^2 dr \left[ -\alpha^X_{\ell_1}(r) \beta^Y_{\ell_2}(r) \beta^Z_{\ell_3}(r) 
+ 2 \;{\rm perm.}
+ \beta^X_{\ell_1}(r) \gamma^Y_{\ell_2}(r) \delta^Z_{\ell_3}(r) + 5 \;{\rm perm.}
- 2 \delta^X_{\ell_1}(r) \delta^Y_{\ell_2}(r) \delta^Z_{\ell_3}(r) 
\right].
\eea
Finally, the bispectrum for orthogonal shape can be cast in the form ~\cite{Senatore:2009gt}
\bea
b^{XYZ}_{\ell_1 \ell_2 \ell_3} & = & ~18 \int\!\!r^2 dr \left[ -\alpha^X_{\ell_1}(r) \beta^Y_{\ell_2}(r) \beta^Z_{\ell_3}(r) 
+ 2 \;{\rm perm.}
+ \beta^X_{\ell_1}(r) \gamma^Y_{\ell_2}(r) \delta^X_{\ell_3}(r) + 5 \;{\rm perm.}
- \frac{2}{3} \delta^X_{\ell_1}(r) \delta^Y_{\ell_2}(r) \delta^Z_{\ell_3}(r) 
\right] \,. 
\eea

{\it Flat Sky Limit:} For simplicity we adopt the flat sky approximation throughout. 
The anisotropies can be decomposed in spin-0 and spin $\pm 2$ plane waves in Fourier space
\begin{eqnarray}
\Delta T(\bfx)&=&\int \frac{d^2\bfk}{(2\pi)^2} a^T(\bfk)e^{i\bfk\cdot \bfx} \nonumber \\ 
\left[Q\pm iU\right] (\bfx)&=& \int  \frac{d^2 \bfk}{(2\pi)^2} \left[ a^E(\bfk)\pm ia^B(\bfk)\right] e^{2i\varphi_k}e^{i\bfk\cdot \bfx} \,.
\end{eqnarray}
The Fourier transform of 3-point function is
\be
\langle a^X(\bfk_1)a^Y(\bfk_2)a^Z(\bfk_3) \rangle =(2\pi)^2B^{XYZ}(k_1,k_2,k_3)\delta^2(\bfk_1+\bfk_2+\bfk_3)
\ee
and the bispectrum $B(k_1,k_2,k_3)$ simplifies to
\be
B^{XYZ}(k_1,k_2,k_3)\approx  b^{XYZ}_{\ell_1 \ell_2 \ell_3}\,,
\ee
where $\vert \bfk_i\vert=\ell_i$.

The bispectrum signal-to-noise is 
\be
\frac{S}{N}= \frac{1}{\sqrt{{\mathcal F}^{-1}}}\,,
\ee
where the Fisher matrix ${\mathcal F}$ is given by ~\cite{KS2001,BZ04,Yadav_etal08a,YKW07}: 
\be
{\mathcal F} = \sum_{X'Y'Z'}\sum_{XYZ}\int {d^2 \ell_1 \over (2\pi)^2}  {d^2 \ell_1 \over (2\pi)^2}  {d^2 \ell_1 \over (2\pi)^2}\,\,
 B^{XYZ}_{\ellt} \Big( \tilde C_{\ell_1}^{-1}\Big)^{XX'} \Big(\tilde C^{-1}_{\ell_2}\Big)^{YY'} \Big(\tilde C^{-1}_{\ell_3}\Big)^{ZZ'} B^{X'Y'Z'}_{\ellt}  \,.
\label{eq:fisher}
\ee
The sum over $XYZ$ indicate permutation of $TEE$ and $TTE$, and $(\tilde C_{\ell}^{-1})^{XY}$  are the $XY$ elements of the
matrix                    $\left(                   \begin{array}{ccc}
C_{\ell}^{TT}&C_{\ell}^{TE}\\C_{\ell}^{TE}&C_{\ell}^{EE}\\\end{array} \right)^{-1}\,  $.
We assume all the cosmological parameters, except $\fnl$, are known.  Indices $XYZ$  and $X'Y'Z'$ run over all the eight possible ordered combinations of temperature and polarization given by $TTT$, $TTE$, $TET$, $ETT$, $TEE$, $ETE$, $EET$ and $EEE$.
The overlap between any two bispectrum shape templates can be represented by a 2-d cosine as~\cite{BZ04}
\begin{equation}
\cos(B_1, B_2)=\frac{B_1 \cdot B_2}{(B_1 \cdot B_1)^{1/2}(B_2 \cdot B_2)^{1/2}}
\label{eqn:2dcosine}
\end{equation}
where the dot product is defined as
\begin{eqnarray}
B_1 \cdot B_2 = \int \frac{d^2 \ell_1}{(2\pi)^2} \frac{d^2 \ell_2}{(2\pi)^2} \frac{d^2 \ell_3}{(2\pi)^2}   B_1(\ell_1,\ell_2, \ell_3)B_2(\ell_1,\ell_2, \ell_3)/(C_{\ell_1} C_{\ell_2} C_{\ell_3}).
\end{eqnarray}
The maximum magnitude of 2d-cosine, 1, correspond to a complete overlap between the two templates. Two completely orthogonal shapes will give rise to zero 2d-cosine.

\section{Instrumental Systematics}
\label{sec:systematics}

The detection and measurement of CMB temperature anisotropy have already provided compelling evidence that the primordial perturbations have been generated during an inflationary period in the very early Universe. The next challenge is to constrain inflationary models by means of precise measurement of CMB temperature and polarization anisotropies. As we have noted, the bispectrum of CMB is a powerful probe of the early universe. However, the instrumental systematic effects on the CMB bispectrum are relatively unexplored. 

For temperature systematics we can introduce the linear calibration (gain fluctuation) parameter of receivers, $a(\bn)$.
The non-linear response of the detectors to observed CMB fields parameter is parametrized by $b(\bn)$.
Effects associated with optical imperfections, i.e. beam shape, 
are collectively captured by the parameter $c$
\begin{equation}
\cmb^{obs}(\bn) =
[1+a(\bn)]\cmb(\bn)+b(\bn)\cmb(\bn)^2+[c\otimes\cmb](\bn)+...\,,
\label{eqn:Tsys}
\end{equation}
where $\cmb^{obs}(\bn)=\Delta T(\bn)/T$ is observed CMB temperature fluctuation at direction $\bn$ including detector noise. We will see the effect of these systematics on the temperature bispectrum and eventually on $\fnl$ in the next section. The last term accounts for effects associated with beam systematics (or more generally, any effect that 
can be described as a convolution with the underlying sky in {\it real space}). Here $\otimes$ denotes 2D real-space convolution. One immediate consequence of this is that by the convolution theorem this will correspond to a product in multipole space, the last term then reads 
$\tilde{c}({\bf l})\tilde{\cmb}({\bf l})$ which implies no $l-l$ coupling in multipole space. Note that this argument neglects the effect of scanning strategy, which we assume is small throughout. This is well-motivated for temperature anisotropy measurements. However, this is not necessarily the case when the (very weak) $B$-mode polarization is considered and beam imperfections (such as differential pointing or differential beamwidth) leak the much larger temperature anisotropy to $B$-mode polarization via a scanning-strategy-dependent {\it local} weight which enters the $B$-mode map-making process (as well as maps of $T$ and $E$-mode) and generally varies across the sky and depends solely on the non-uniform scanning strategy. Since only little information is contained in the $B$-mode polarization on primordial non-Gaussianity we will ignore this effect here. In principle, leakage of temperature anisotropy to $E$-mode polarization cannot be ignored and indeed we give a special attention to these terms here. However, we ignore the impact of the coupling of non-uniform scanning strategy to beam distortions on the spurious $E$-mode polarization and its propagation to $\fnl$ bias as this is a higher order effect and highly depends on the details of scanning strategy.

{\it Polarization systematics:} The measurement of CMB polarization is complicated by its low level signal, galactic foregrounds, and various systematic effects. In particular, precise control of systematics is required, especially for polarization measurements. Measuring the polarization using Polarization Sensitive Bolometer (PSB) pairs inside a feedhorn, each measuring the intensity of one direction of the polarization, is used by several ground- and balloon-based CMB experiments such as \emph{Planck}\footnote{\texttt{http://www.rssd.esa.int/index.php?project=Planck}}, \emph{MAXIMA}\footnote{\texttt{http://cosmology.berkeley.edu/group/cmb/}}, \emph{Boomerang}\footnote{\texttt{http://cmb.phys.cwru.edu/boomerang/}}, \emph{BICEP}\footnote{\texttt{http://bicep.caltech.edu/}}, \emph{EBEX}\footnote{\texttt{http://groups.physics.umn.edu/cosmology/ebex/}} etc. The difference of the received intensity from the two PSB gives a combination of $Q$ or $U$ in the frame of the focal plane. Polarization measurements therefore involve taking the difference of received intensity at two orthogonal polarizers. Any differences between the two PSBs might generate spurious $Q$ and $U$ signals. Furthermore, the beams of the horns generally have some degree of ellipticity. The typical differential ellipticity of a pair is around few percent, while for a single horn is usually at sub-percent level $\sim 0.5\%$~\cite{Rosset2007}. When several detectors are combined to obtain $Q$ and $U$ maps, it is important to match the responses of these detectors precisely in terms of cross-calibration, beam shape, spectral response, etc. Below we discuss the parametrization of CMB polarization systematics. 

The polarization systematics fall into two categories, one associated with the detector system which distorts the
polarization state of the incoming polarized signal (Type I hereafter), and another associated
with systematics of the CMB signal due to the beam anisotropy (Type II hereafter). Each type of systematics contamination can be further classified into linear and non-linear effects. We parametrize the linear instrumental systematics for CMB polarization measurements following~\cite{HHZ} (HHZ from hereafter), and introduce corresponding analogous non-linear systematics parametrization. In Appendix A we briefly discuss an alternative parametrization employing a Stokes-parameter-based formalism and show the equivalence of the two parametrizations.

The instrumental response to incoming CMB radiation is usually described by the Jones
transfer matrix. Bias induced in the matrix determination will mix the corresponding Stokes parameters. 
To first order, the effect of Type I systematics on the Stokes parameters can be written as~\cite{HHZ}
\begin{equation}
\delta [Q \pm i U](\bn) =
            [\calb \pm i 2 \rot](\bn)  [Q \pm i U](\bn) + [f_1 \pm i f_2](\bn)   [Q \mp i U](\bn)
            + [\gamma_1 \pm i \gamma_2](\bn) \cmb(\bn)\,,
\label{eq:pointsys}
\end{equation}
where $a(\bn)$ is a scalar field which describes the miss-calibration of the polarization measurements,
$\rot(\bn)$ is another scalar field that describes the rotation miss-alignment of the instrument,
$(f_1\pm if_2)(\bn)$ are spin $\pm 4$ fields that describe the coupling between two spin $\pm 2$
states (spin-flip), and $(\gamma_1\pm i \gamma_2)(\bn)$ are spin $\pm2$ fields that describe
monopole leakage, i.e. leakage from temperature anisotropy to polarization. 

Similar to the Type I systematics, the effect of Type II systematics on
the Stokes parameters can be written as~\cite{HHZ}
\begin{equation}
\label{eq:localmodel}
\delta[Q \pm i U](\bn;\sigma) = \sigma {\bf p}(\bn) \cdot \nabla [Q \pm i U](\bn;\sigma)
+ \sigma [d_1 \pm i d_2](\bn) [\partial_1 \pm i\partial_2] \cmb(\bn;\sigma)
 + \sigma^2 q(\bn) [\partial_1 \pm i \partial_2]^2 \cmb(\bn;\sigma)\, ,
\end{equation}
where the systematic fields are smoothed over the average beam $\sigma$ of the
experiment.  Therefore, the type II systematic fields are sensitive to the 
imperfection of the beam on the beam scale, $\sigma$.
The spin $\pm 1$ fields, $(p_1\pm ip_2)$ and $(d_1\pm id_2)$, describe pointing errors
and dipole leakage from temperature to polarization, respectively, 
and $q$ is a scalar field that represents quadrupole leakage~\cite{HHZ}, e.g. beam ellipticity.

We parametrize the non-linearities of the instrument as
\begin{equation}
\delta [Q \pm i U](\bn) =  [\tilde \gamma_1 \pm i \tilde \gamma_2](\bn) \cmb(\bn)^2 + \sigma [\tilde
d_1 \pm i \tilde d_2](\bn) [\partial_1 \pm i\partial_2]
\cmb^2(\bn;\sigma) + \sigma^2 \tilde q(\bn) [\partial_1 \pm i
\partial_2]^2 \cmb^2(\bn;\sigma)+  ... \,,
\label{eq:nonlinearpointsys}
\end{equation}
where $\tilde \gamma_1$ and $\tilde \gamma_2$ are spin $\pm2$ non-linear parameters which describe the second-order leakage from temperature to polarization states; $\tilde d_1$ and $\tilde d_2$ are spin $\pm1$ parameters which describe the dipole leakage from second order temperature to
polarization states. $\tilde q$ is a scalar field that represents
quadrupole leakage from second order temperature. In general, we use
$\tilde S$ for systematics of non-linear response to distinguish the
same type $S$ linear systematics. Note that in principle there can be terms proportional to $(Q \pm iU)^2(\bn)$, etc, which we have dropped as the $\cmb^2(\bn)$ terms provide the dominant non-linear contribution.

\section{Systematics-Induced CMB Temperature Bispectrum}
\label{sec:distortions}

In this section, we discuss the impact of instrumental systematics on the {\it measured} bispectrum. We show that the linear calibration systematics of detectors can only distort the primordial CMB bispectrum; any detection with only the presence of linear systematics implies the detection of primordial non-Gaussianity. However, non-linear systematics can generate new spurious bispectrum even in the absence of vanishing underlying bispectrum, i.e. the primordial CMB is purely Gaussian.
As discussed before, {\it beam shape effects} will only modify the $\ell$-dependence of an already existing bispectrum, i.e. in case that $\fnl=0$ these beam effects are irrelevant.

With both linear $a(\bn)$ and non-linear $b(\bn)$ systematics included, as in Eq.~(\ref{eqn:Tsys}), the CMB 3-point function of the temperature can be written as
\begin{eqnarray}
\langle\cmb^{obs}(\bn_1)\cmb^{obs}(\bn_2)\cmb^{obs}(\bn_3)\rangle_{CMB,syst} &=& \langle\cmb(\bn_1)\cmb(\bn_2)\cmb(\bn_3)\rangle_{CMB}+ \langle a(\bn_1)\rangle_{syst}\langle\cmb(\bn_1)\cmb(\bn_2)\cmb(\bn_3)\rangle_{CMB}+perm \nonumber\\
&& \quad + \langle b(\bn_1)\rangle_{syst}\langle\cmb^2(\bn_1)\cmb(\bn_2)\cmb(\bn_3)\rangle_{CMB}+perm\nonumber\\
&& \quad + \langle a(\bn_1)a(\bn_2)\rangle_{syst}\langle\cmb(\bn_1)\cmb(\bn_2)\cmb(\bn_3)\rangle_{CMB}+perm \nonumber\\
&& \quad + \langle a(\bn_1)b(\bn_2)\rangle_{syst}\langle\cmb(\bn_1)\cmb^2(\bn_2)\cmb(\bn_3)\rangle_{CMB}+perm \nonumber\\
&& \quad + \langle a(\bn_1)a(\bn_2)a(\bn_3)\rangle_{syst}\langle\cmb(\bn_1)\cmb(\bn_2)\cmb(\bn_3)\rangle_{CMB}\nonumber\\
&& \quad + \langle b(\bn_1)b(\bn_2)\rangle_{syst}\langle\cmb^2(\bn_1)\cmb^2(\bn_2)\cmb(\bn_3)\rangle_{CMB}+perm \,,
\label{E:expandsys}
\end{eqnarray}
where $\langle...\rangle_{CMB}$ means averaging over CMB realizations, and $\langle...\rangle_{syst}$ stands for averaging over systematics field realizations. In principle, there can be higher order terms proportional to $\cmb^3(\bn)$. For 3-point correlation functions, we are only interested in terms up to second order in CMB fields. However, for the effects on trispectrum estimations, 3rd order terms should be considered. The first term on the right hand side of Eq.~(\ref{E:expandsys}) is the cosmological 3-point correlation function, i.e. the target signal. The second term is proportional to linear systematics field, e.g. first order gain response. If there is non-zero monopole gain systematics, $\langle a(\bn)\rangle_{syst}$, then this term  will induce a linear bias proportional to the primordial bispectrum. For the third term, if there is any non-zero $\langle b(\bn)\rangle_{syst}$, then this term will introduce a {\it new bispectrum} even if there is no primordial non-Gaussianity. The forth term contains self correlations of linear distortion field parameter and is capable of only distorting existing non-Gaussianity. In contrast, the fifth term coming from cross-correlation between linear and non-linear systematics, can generates new bispectrum. If the systematics field is non-Gaussian then the term containing three distortion parameters (the term proportional to $\langle a(\bn_1)a(\bn_2)a(\bn_3)\rangle_{syst}$) can also contaminate $\fnl$ measurements. We do not consider this term and terms involving $\langle b(\bn_1)b(\bn_2)\rangle_{syst}$ in this paper.

In case that a sufficiently small sky patch is considered, spherical harmonic modes can be replaced by Fourier modes. Generalization from the flat-sky to the full sky is straightforward and is given in Appendix B. If we consider the non-linear calibration up to second order, the Fourier transform of the Taylor-expanded observed temperature anisotropy (including calibration contamination) could be written as
\begin{eqnarray}
\cmb(\bfl)^{obs}
&=&\int d \bn\, \cmb^{obs}(\bn) e^{-i \bfl \cdot \bn}
= \cmb(\bfl) - \intlnp \cmb(\bflp) \Bigg[a(\bfl-\bflp) +  \intlnpp  \cmb(\bflpp)b(\bfl - \bflp -
\bflpp) + \ldots \Bigg]\,.
\end{eqnarray}
Using the flat-sky approximation, we can define the angular bispectrum as
\begin{equation}
\langle \cmb(\bfl_1) \cmb(\bfl_2) \cmb(\bfl_3)\rangle \equiv (2\pi)^2
\delta(\bfl_1+\bfl_2+\bfl_3) B^{\cmb\cmb\cmb}_{(\bfl_1,\bfl_2,\bfl_3)}\,.
\end{equation}

The effect of systematics on bispectrum can be obtained by Fourier transforming Eq.~(\ref{E:expandsys}). For simplicity we will classify the effects from systematics into two categories; one where the systematics create new bispectrum even if the underlying CMB is purely Gaussian, referred to as a `bias term', and second, where the shape of the underlying bispectrum is distorted due to systematics. The first contribution can be written as
\begin{eqnarray}
B^{\cmb\cmb\cmb,bias}_{(\bfl_1,\bfl_2,\bfl_3)} =
\bigg[ C_{\ell_2}^{a\cmb} C_{\ell_3}^{\cmb\cmb} + \text{5 perm.}\bigg] + \bigg[b(0)C^{\cmb\cmb}_{\ell_1}C^{\cmb\cmb}_{\ell_2} + \text{5 perm.}\bigg]  + \intlp \bigg[C^{ab}_{\bfl_1-\bfl'}C^{\cmb\cmb}_{\ell^{'}}C^{\cmb\cmb}_{\ell_3} + \text{5 perm.}\bigg].
\label{eq:tempgeneration}
\end{eqnarray}
The first term is zero since one does not expect systematics to be correlated with the CMB. This term is analogous to the integrated Sachs-Wolfe (ISW) effect which cross correlate the lensing potential with CMB field. The second term is proportional to $b(0)$ which is the bispectrum generated from the monopole of non-linear systematics field. The terms involving the cross correlation between the $a(\bfl)$ and $b(\bfl)$ fields can induce a bispectrum as well. We note that with no cross-correlation between the linear $a(\bfl)$ and non-linear $b(\bfl)$ field, only the monopole of the non-linear parameter $b(0)$ would bias the bispectrum detection. 
The term describing the systematic distortion of primordial bispectrum can be written as
\begin{eqnarray}
&&\delta  B^{\cmb\cmb\cmb,dist}_{(\bfl_1,\bfl_2,\bfl_3)} = a(0)B^{\cmb\cmb\cmb}_{(\bfl_1,\bfl_2,\bfl_3)} + \intlp C^{aa}_{\ell^{'}}
\Big[B^{\cmb\cmb\cmb}_{(\bfl_1,\bfl_2-\bfl',\bfl_3+\bfl')} 
 + B^{\cmb\cmb\cmb}_{(\bfl_1-\bfl',\bfl_2+\bfl',\bfl_3)} + B^{\cmb\cmb\cmb}_{(\bfl_1+\bfl',\bfl_2,\bfl_3-\bfl')}\Big] \,.
\label{eq:tempdistortion}
\end{eqnarray}
The total observed bispectrum $B^{\cmb\cmb\cmb,obs}_{(\bfl_1,\bfl_2,\bfl_3)}$ would then be
\begin{equation}
 B^{\cmb\cmb\cmb,obs}_{(\bfl_1,\bfl_2,\bfl_3)} =B^{\cmb\cmb\cmb,primordial}_{(\bfl_1,\bfl_2,\bfl_3)} +\delta B^{\cmb\cmb\cmb,dist}_{(\bfl_1,\bfl_2,\bfl_3)}  +B^{\cmb\cmb\cmb,bias}_{(\bfl_1,\bfl_2,\bfl_3)}\,.
\end{equation}

\section{Systematics-Induced CMB Polarization Bispectrum}
\label{sec:polabispectrum}

Currently  the best CMB constraints on  the three parameters $f^{local}_{NL}, f^{equil.}_{NL}, f^{ortho}_{NL}$ come from the WMAP
temperature anisotropy data~\cite{Creminelli_wmap2,YW08,Smith:2009jr,Senatore:2009gt,Komatsu:2010fb}. By also having the polarization
information, one can improve
sensitivity to  primordial fluctuations \cite{BZ04,YW05,Yadav_etal08a,Yadav2007}. The ongoing Planck experiment and futuristic CMB experiments such as CMBPol will characterize the polarization anisotropy to high
accuracy. In this section we discuss the instrumental polarization systematics which can contaminate the primordial non-Gaussian signal.

\begin{table}
\begin{tabular}{c||c}
\hline
\bf {Linear Systematics $S$}& $W^S_{E}(\bfl_1,\bfl_2)$ \\
\hline
Calibration $a$ &  $\cos[2 (\varphi_{\bfl_2} - \varphi_\bfL)]$\\
Rotation $\rot$ &  $-2 \sin[2 (\varphi_{\bfl_2} - \varphi_\bfL)]$ \\
Pointing $p_a$   & $\sigma (\vl_2  \cdot \hat \vl_1) \sin[ 2(\varphi_{\bfl_2} - \varphi_\bfL)]$ \\
Pointing $p_b$  & $-\sigma (\vl_2 \times \hat \vl_1)\cdot \hat{\bf z} \sin[ 2(\varphi_{\bfl_2}- \varphi_\bfL) ]$\\
Flip $f_a$   & $\cos[2 (2 \varphi_{\bfl_1} - \varphi_{\bfl_2} -\varphi_{\bfL}) ]$\\
Flip $f_b$   & $-\sin[2 (2 \varphi_{\bfl_1}-\varphi_{\bfl_2}-\varphi_{\bfL})]$ \\
Monopole $\gamma_a$ &$\cos[2 (\phi_{\bfl_1}- \varphi_\bfL)]$ \\
Monopole $\gamma_b$ & $-\sin[2 (\varphi_{\bfl_1}- \phi_\bfL)$ \\
Dipole $d_a$      & $(l_2 \sigma) \sin[ \varphi_{\bfl_1} + \phi_{\bfl_2} - 2 \varphi_\bfL]$ \\
Dipole $d_b$      & $ (l_2 \sigma) \cos[ \varphi_{\bfl_1} + \varphi_{\bfl_2} - 2 \varphi_\bfL]$ \\
Quadrupole $q$       &$ - (l_2 \sigma)^2 \cos[ 2 (\varphi_{\bfl_2} - \varphi_\bfL)]$ \\
\hline
\end{tabular}
\caption{Window functions for all the 11 linear response systematic parameters.
First column indicates the type of systematic parameter in
consideration. Second and third columns show the window
functions needed to calculate systematic effects on bispectrum measurement and appear in Eq.~(\ref{eqn:Poldistortion}). 
We note that ${\vl_1} = l_1 \bl_1,\,\vl_2 = \bfL - \vl_1$, and ${\vl_2} = l_2 \bl_2$ and $\varphi_{\bf l}=\cos^{-1}(\hat {\bf n } \cdot \hat {\bf l})$.}
\label{table:linear}
\end{table}

\subsection*{Impact of linear response on polarization bispectrum}
\label{sec:appentemsys}

Here we show the calculation for spurious bispectrum
distorted by linear instrumental systematics involving
polarization field. Performing the harmonic transformation of Eq.~(\ref{eq:pointsys}) and Eq.~(\ref{eq:localmodel}),
we get the distorted polarization $E$-mode from linear instrumental systematics
\begin{eqnarray}
E(\vl)^{obs} &=& \int d \bn\, [\tQ^{obs}(\bn) \cos(2\varphi_\vl) - \tU^{obs}(\bn) \sin(2\varphi_\vl)] e^{-i \bfl \cdot \bn}. \nonumber\\
\label{E:thetal}
\end{eqnarray}
We can easily calculate the bispectrum involving observed $E$-mode to find out the changes induced by different types of linear instrumental systematics
\begin{equation}
\langle X^{obs}(\bfl_1) Y^{obs}(\bfl_2) Z^{obs}(\bfl_3)\rangle \equiv (2\pi)^2
\delta(\bfl_1+\bfl_2+\bfl_3) B^{XYZ}_{(\bfl_1,\bfl_2,\bfl_3)}\,
\end{equation}
where $X, Y, Z$ is temperature or $E$-mode polarization field of CMB. For example, the bispectrum $\delta B^{EEE,dist}_{(\bfl_1,\bfl_2,\bfl_3)}$ describing the distortion of primordial polarization bispectrum due to instrumental systematics can be written as
\begin{eqnarray}
\delta  B^{EEE,dist}_{(\bfl_1,\bfl_2,\bfl_3)} &=& \sum_{SS'}
\intlp\intlpp C^{SS'}_{\ell^{'}}
\Big[B^{EEE}_{(\bfl_1,\bfl_2-\bfl',\bfl_3+\bfl')}
W^{S}_{E}(\bfl_1,-\bfl')W^{S'}_{E'}(\bfl_2,-\bfl'') \nonumber \\
&& \quad + B^{EEE}_{(\bfl_1-\bfl',\bfl_2+\bfl',\bfl_3)}W^{S}_{E}(\bfl_1,-\bfl')W^{S'}_{E'}(\bfl_3,-\bfl'')
+ B^{EEE}_{(\bfl_1+\bfl',\bfl_2,\bfl_3-\bfl')}W^{S}_{E}(\bfl_3,-\bfl')W^{S'}_{E'}(\bfl_3,-\bfl'') \Big]\,. 
\label{eqn:Poldistortion}
\end{eqnarray}
Summations are over 11 systematics contributions, $S= {a, \rot, f_a, f_b, p_a ,p_b, \gamma_a, \gamma_b, d_a, d_b, q}$. Window functions $W^S_E(\bfl_1,\bfl_2)$ are given Table~\ref{table:linear}.

\subsection*{Impact of non-linear response on polarization bispectrum}
\label{sec:nonlinearpola}

As for the linear response systematics, we can calculate the observed CMB $E$-polarization in the presence of non-linear systematic fields. We focus on the non-linear leakage from second order temperature to $E$-polarization, as this is the dominant systematics which could potentially bias bispectrum detection.  We will show that the monopole of the non-linear response systematics $\tilde\gamma_1,\tilde\gamma_2, \tilde d_1, \tilde d_2,$ and $\tilde q$ fields can generate new bispectrum. To second order, any cross-correlation between linear and non-linear systematics could also generate non-zero bispectrum. The $E(\bfl)$ field in the presence of the non-linear systematics $\tilde S$ can be simplified as follows
\begin{equation}
E^{obs}(\bfl)=E^{obs}(\bfl)+\int {d^2 \bfl' \over (2\pi)^2}{d^2 \bfl'' \over (2\pi)^2} {\tilde S}(\bfl-\bfl'-\bfl'') \tilde T(\bfl') \tilde T(\bfl'')W^{\tilde S}_{E}(\bfl,\bfl',\bfl'') \,. \nonumber \\
\label{eqn:enb1}
\end{equation}
Unlike the temperature bispectrum case, for polarization there are many terms containing cross-correlation between non-linear systematics ($\tilde \gamma,\tilde d,$ or $\tilde q$) and linear systematics S (11 systematics as given in Table~\ref{table:linear}).  The spurious bispectrum contribution for $EEE$, $TTE$, and the $TEE$ bispectrum can be written as
\begin{eqnarray}
B^{EEE,\tilde S,bias}_{(\bfl_1,\bfl_2,\bfl_3)}&=& \tilde S(0)C^{TE}_{\ell_1}C^{TE}_{\ell_2} W^{\tilde S}_{E}(\bfl_3,-\bfl_1,-\bfl_2) + {\rm perm.}\nonumber \\ 
B^{TTE,\tilde S,bias}_{(\bfl_1,\bfl_2,\bfl_3)}&=& \tilde S(0)C^{TT}_{\ell_1}C^{TT}_{\ell_2}  W^{\tilde S}_{E}(\bfl_3,-\bfl_1,-\bfl_2) + {\rm perm.}\nonumber \\ 
B^{TEE,\tilde S,bias}_{(\bfl_1,\bfl_2,\bfl_3)}&=& \tilde S(0)C^{TT}_{\ell_1}C^{TE}_{\ell_2}  W^{\tilde S}_{E}(\bfl_3,-\bfl_1,-\bfl_2) + \tilde S(0)C^{TT}_{\ell_1}C^{TE}_{\ell_3}  W^{\tilde S}_{E}(\bfl_2,-\bfl_1,-\bfl_3) + {\rm perm.}
\label{eqn:Polsys}
\end{eqnarray}
where $W^{\tilde S}_{E}$ are given in Table~\ref{table:nonlinearsys}.  The total observed bispectrum $B^{XYZ, obs}_{(\bfl_1,\bfl_2,\bfl_3)} $ including both the systematics induced and distorted contribution is
\begin{equation}
B^{XYZ, obs}_{(\bfl_1,\bfl_2,\bfl_3)} =B^{XYZ,primordial}_{(\bfl_1,\bfl_2,\bfl_3)} + \sum_{S}\delta B^{XYZ,dist}_{(\bfl_1,\bfl_2,\bfl_3)} + \sum_{\tilde S}B^{XYZ,bias}_{(\bfl_1,\bfl_2,\bfl_3)} \,.
\end{equation}

\begin{table}
\begin{tabular}{c||c}
\hline
\bf {Non-linear Systematics $\tilde S$} & $W^{\tilde S}_{E}(\bfl,\bflp,\bflpp)$ \\
\hline
Monopole leakage $\tilde\gamma_a$  & $2\cos[2(\varphi_{\bfl-\bfl'-\bfl''}-\varphi_{\bfl})]$\\
Monopole leakage $\tilde\gamma_b$ & $-2 \sin[2(\varphi_{\bfl-\bfl'-\bfl''}-\varphi_{\bfl})]$ \\
Dipole leakage $\tilde d_a$   & $\sigma[\bfl'\sin(\varphi_{\bfl'} + \varphi_{\bfl-\bfl'-\bfl''} - 2 \varphi_\bfl) + \bfl''\sin(\varphi_{\bfl''} + \varphi_{\bfl-\bfl'-\bfl''} - 2 \varphi_\bfl)]$\\
Dipole leakage $\tilde d_b$   & $\sigma[\bfl'\cos(\varphi_{\bfl'} + \varphi_{\bfl-\bfl'-\bfl''} - 2 \varphi_\bfl) + \bfl''\cos(\varphi_{\bfl''} + \varphi_{\bfl-\bfl'-\bfl''} - 2 \varphi_\bfl)]$\\
Quadrupole leakage $\tilde q$  & $-2\sigma^2\{\bfl'^2 \cos[ 2 (\varphi_{\bfl'} - \varphi_\bfl)] + \bfl''^2 \cos[ 2 (\varphi_{\bfl''} - \varphi_\bfl)]\}$\\
\hline
\end{tabular}
\caption{Window functions for 5 non-linear response systematic parameters.  Here ${\vl_1} = l_1 \bl_1,\,\vl_2 = \bfL - \vl_1$, and ${\vl_2} = l_2 \bl_2$, and $\varphi_{\bf l}=\cos^{-1}(\hat {\bf n } \cdot \hat {\bf l})$.}
\label{table:nonlinearsys}
\end{table}

\begin{figure}[t]
\includegraphics[width=55mm,angle=-90]{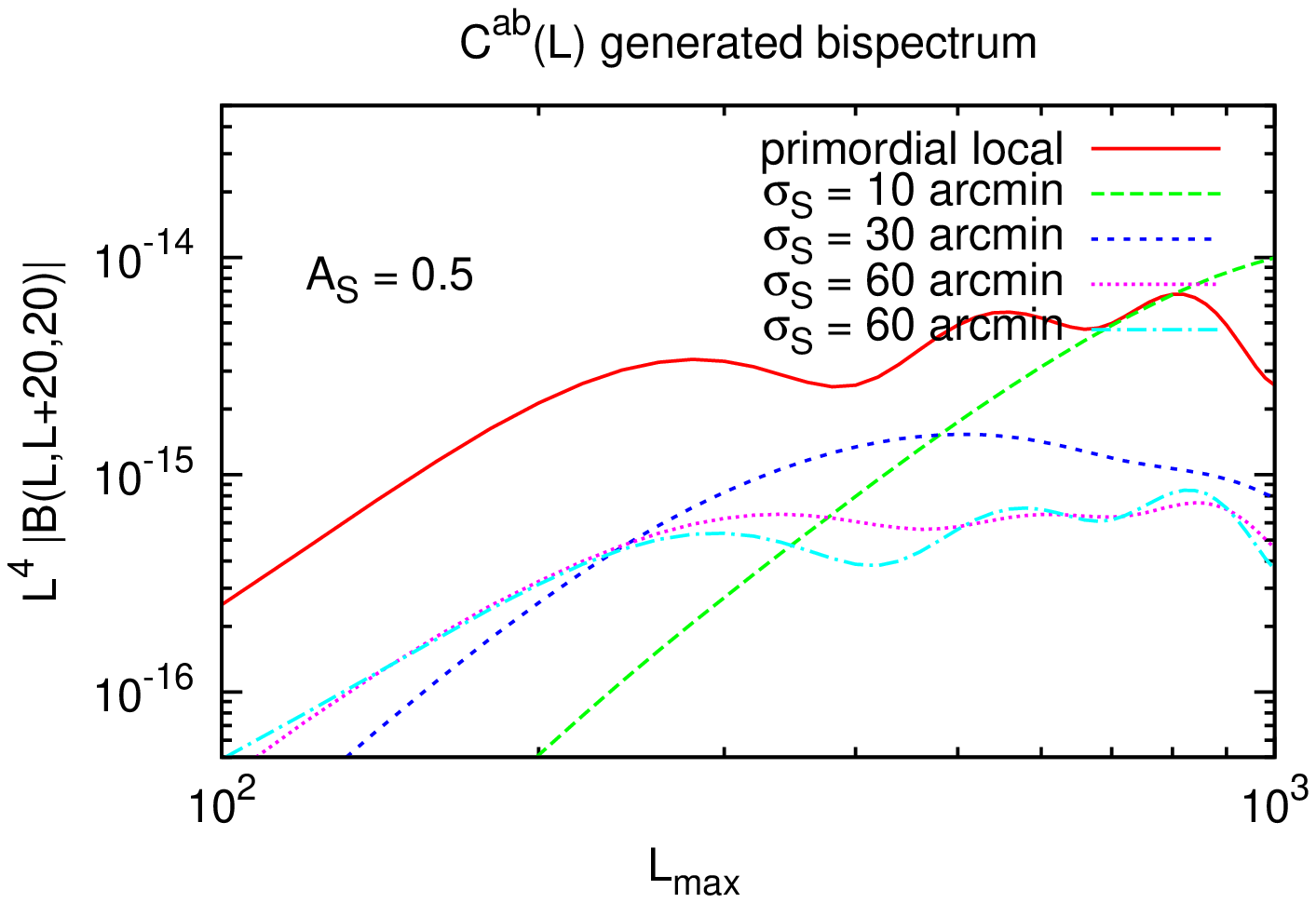}
\includegraphics[width=55mm,angle=-90]{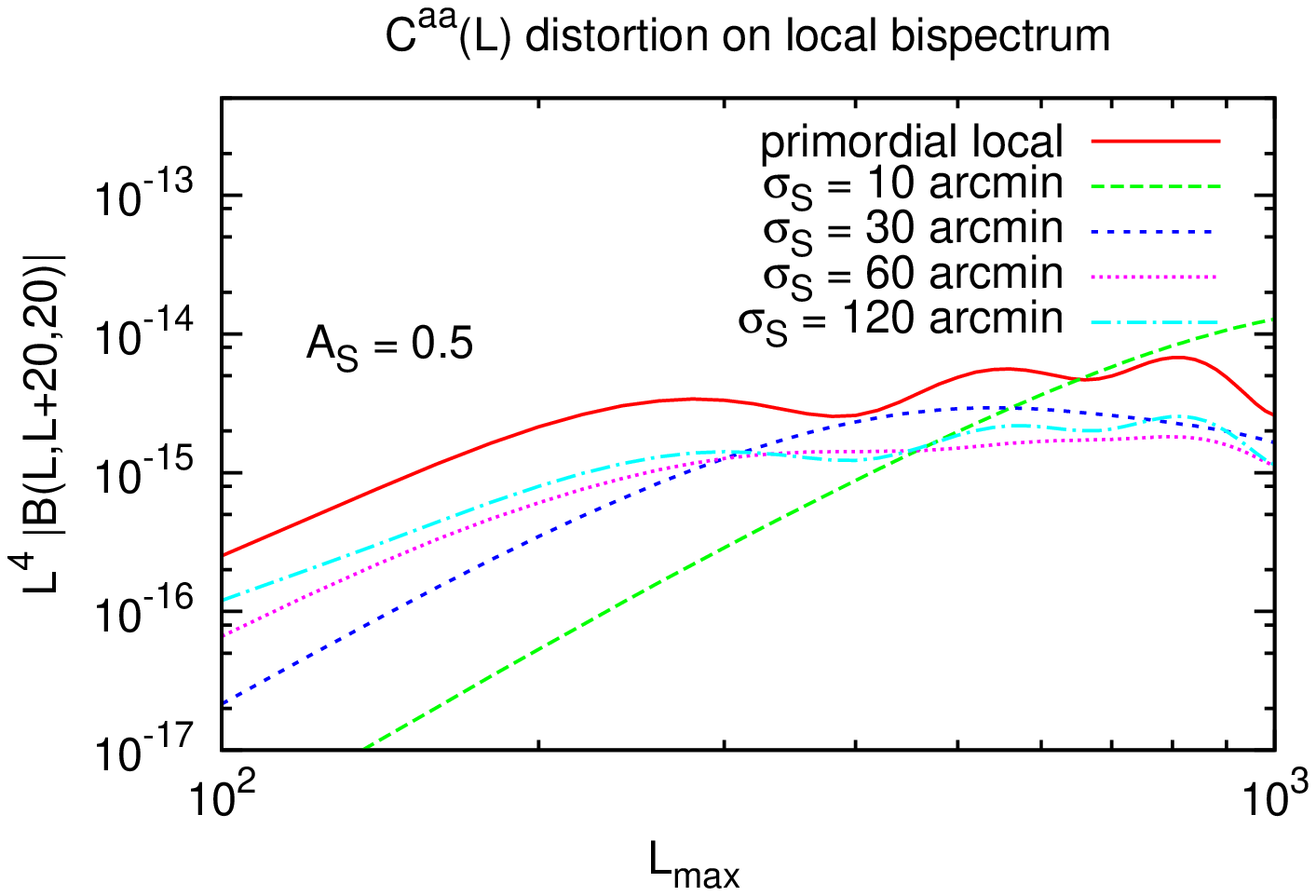}
\includegraphics[width=55mm,angle=-90]{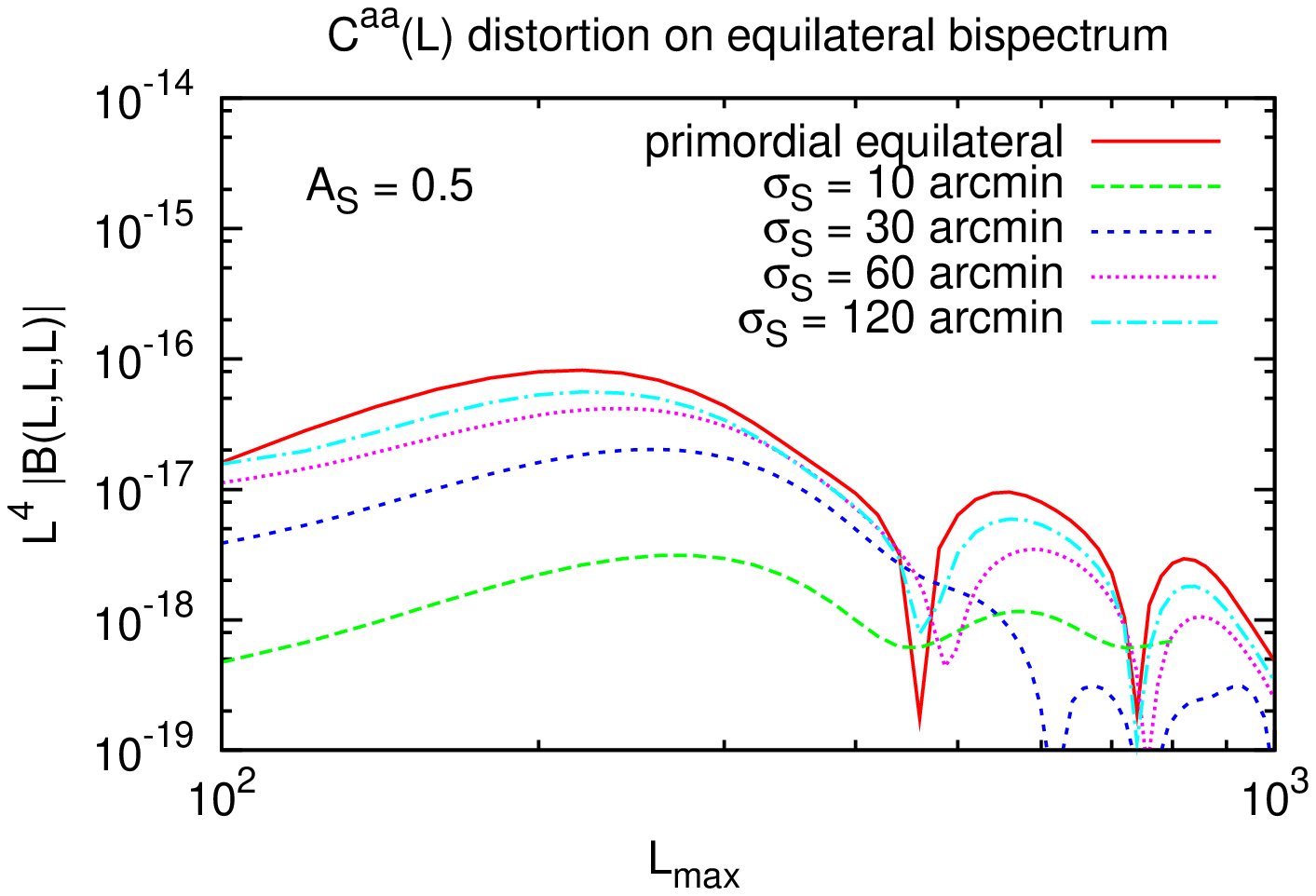}
\includegraphics[width=55mm,angle=-90]{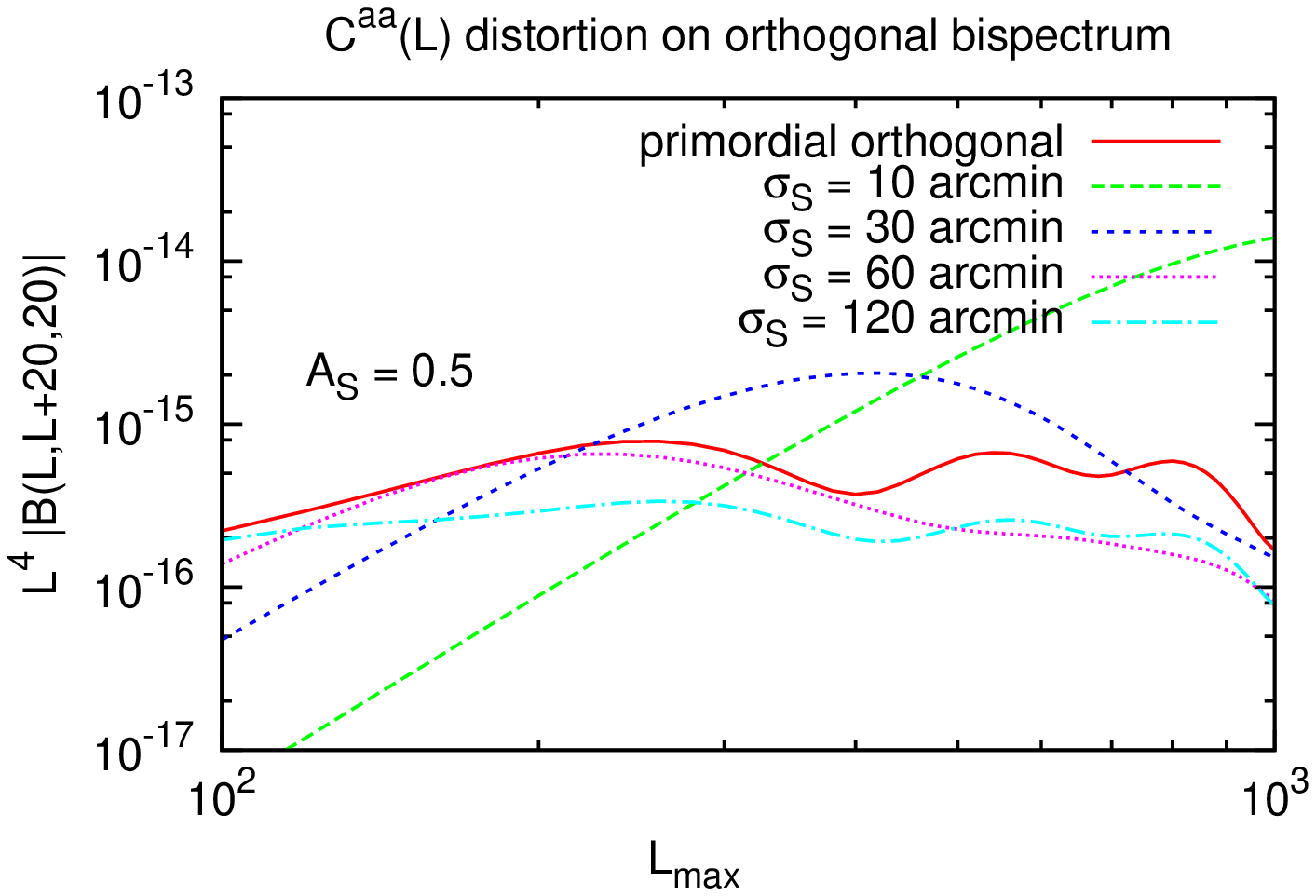}
\caption{{\it Upper left panel:} Spurious bispectrum generated from the cross correlation of calibration parameter $a(L)$ and non-linear response parameter $b(L)$. For reference the solid red curve shows the primordial bispectrum of local type with non-linearity parameter $\fnl=1$. {\it Upper right panel:} Distortions on temperature local type bispectrum from linear instrumental gain fluctuations. For reference the solid red line shows the CMB bispectrum from local type model with $\fnl=1$. {\it Lower left panel:} Similar for {\it upper right panel} but for equilateral type of bispectrum.  {\it Lower right panel:} Similar for {\it upper right panel} but for orthogonal type of bispectrum. In all the panels  We have assumed systematics model as defined in Eq.~(\ref{eq:coh}) and choose the {\it rms} of systematics field $A_a=A_b= 0.5$, while varying coherence length starting from $\sigma_S=10'$ to $\sigma_S=120'$. Note that the spurious signals scale as $A_{S}^{2}$.} 
\label{fig1} 
\end{figure}

\begin{figure}[]t
\includegraphics[width=40mm,angle=-90]{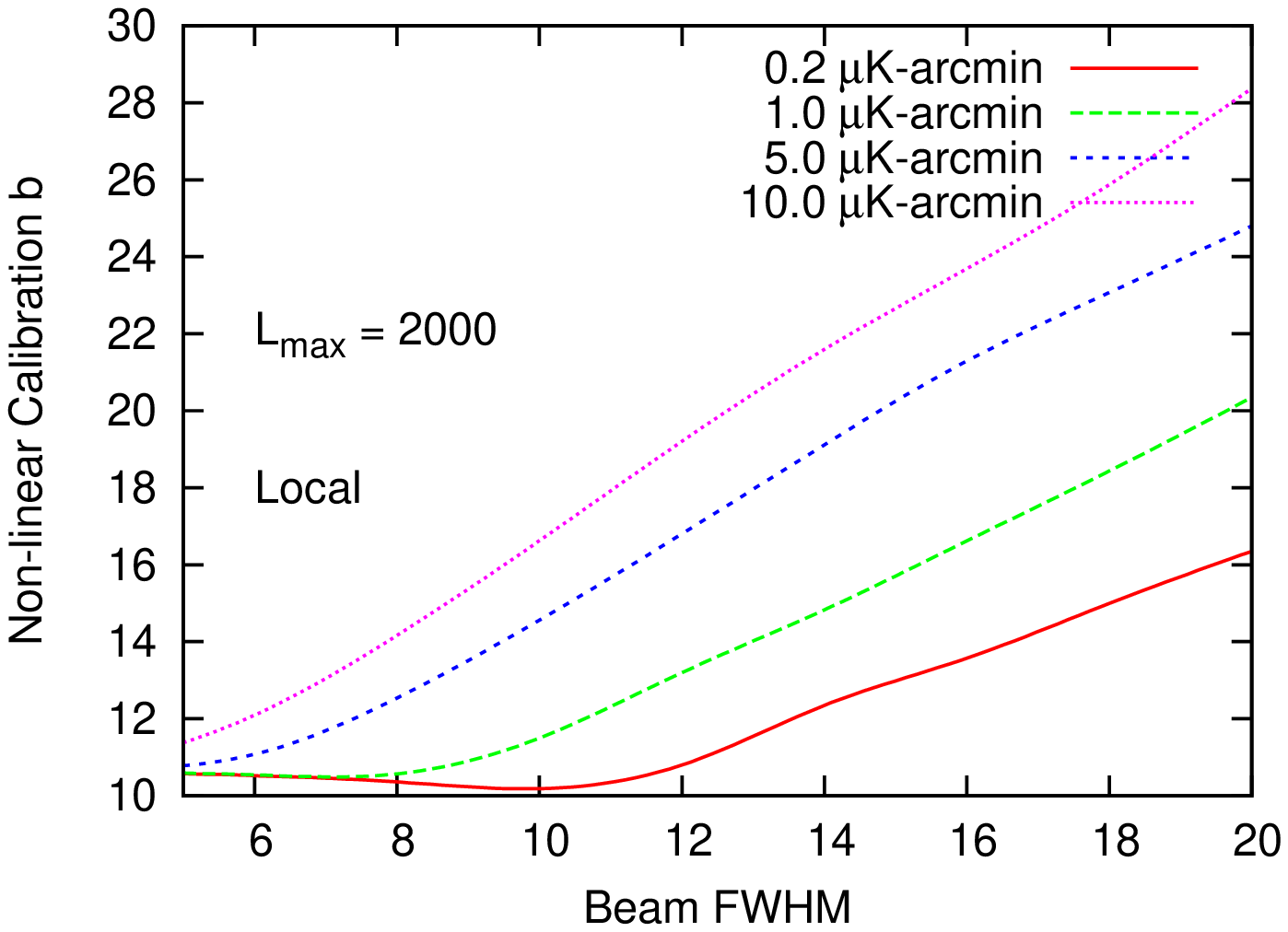}
\includegraphics[width=40mm,angle=-90]{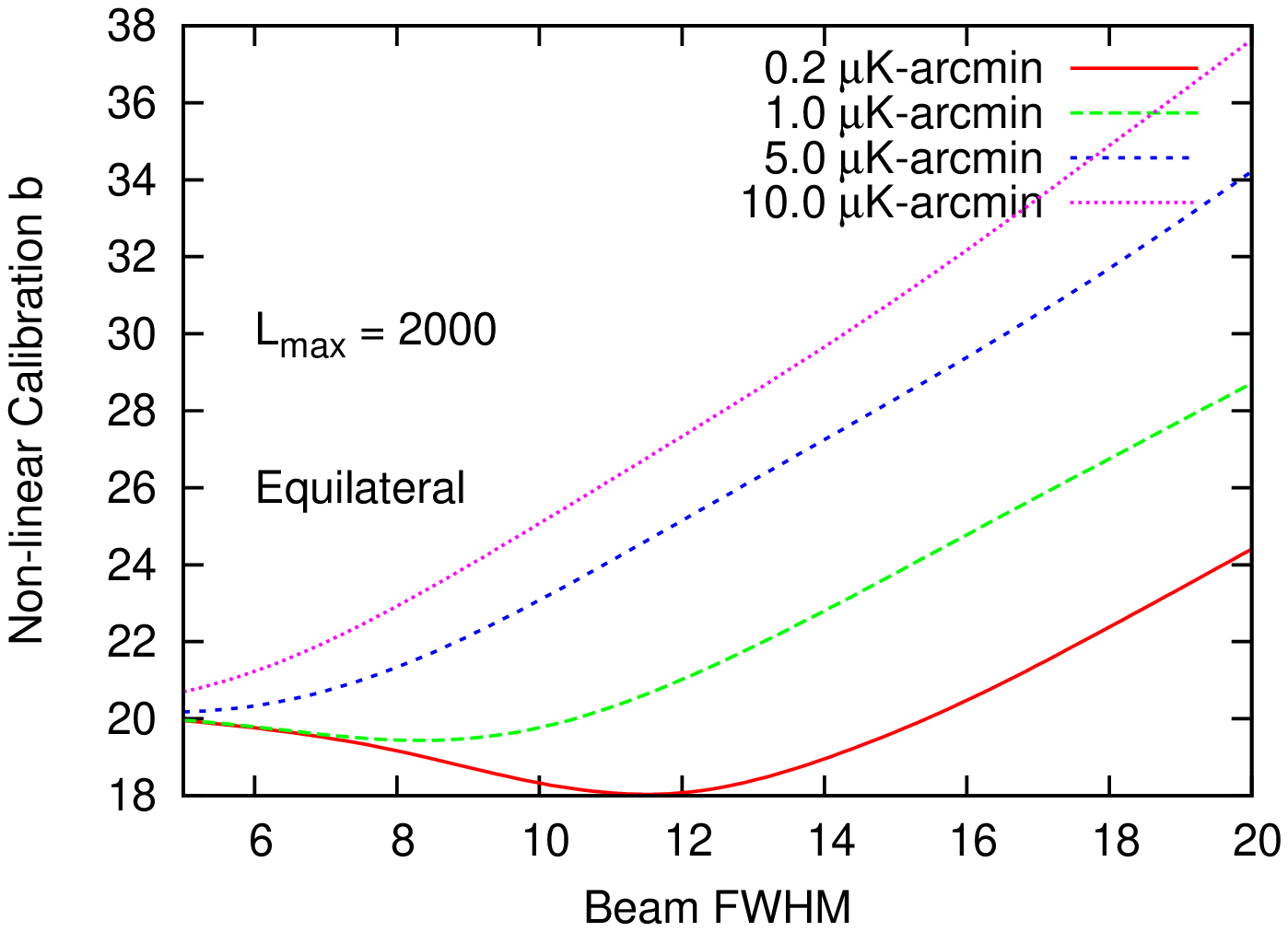}
\includegraphics[width=40mm,angle=-90]{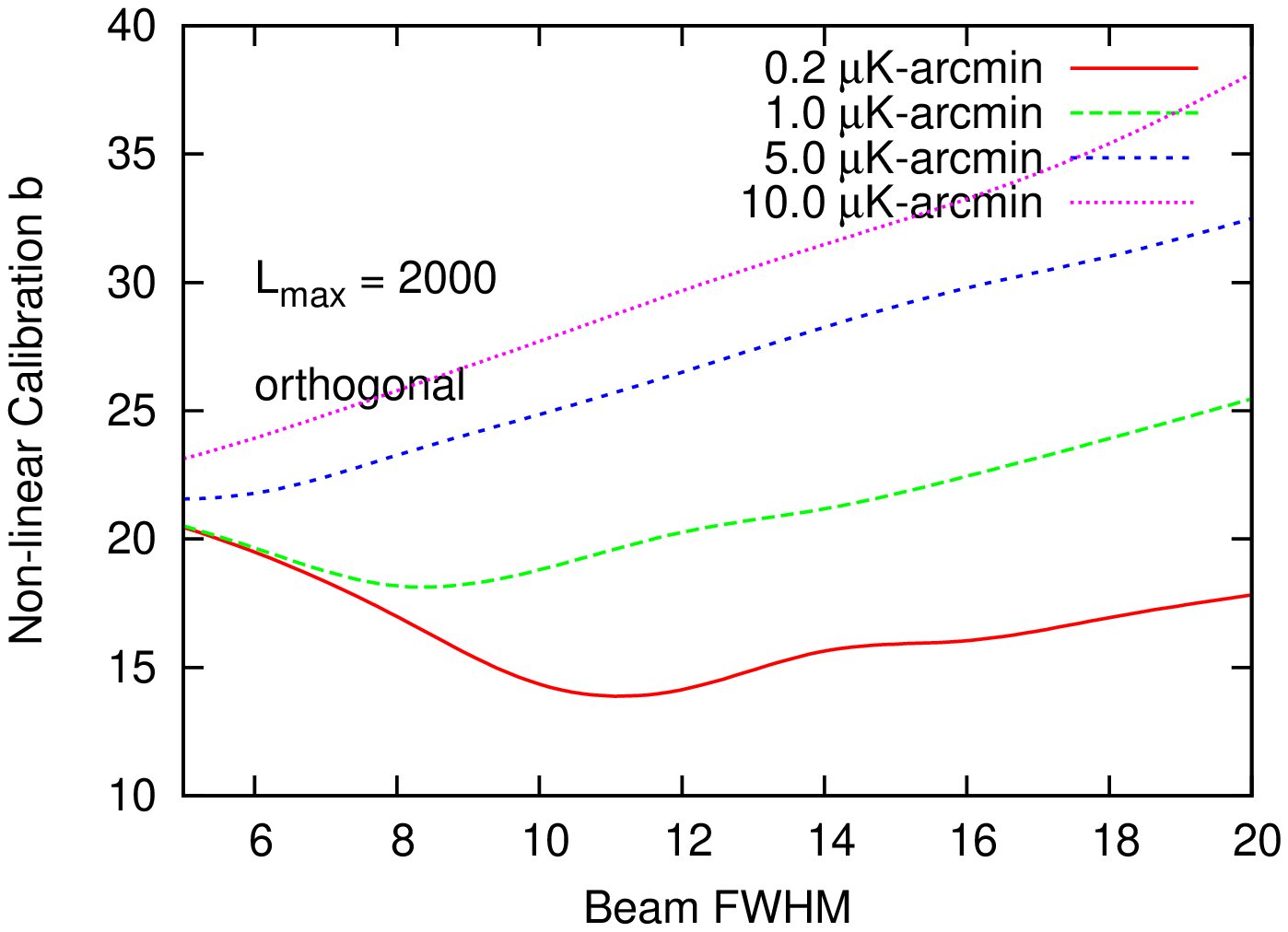}
\includegraphics[width=40mm,angle=-90]{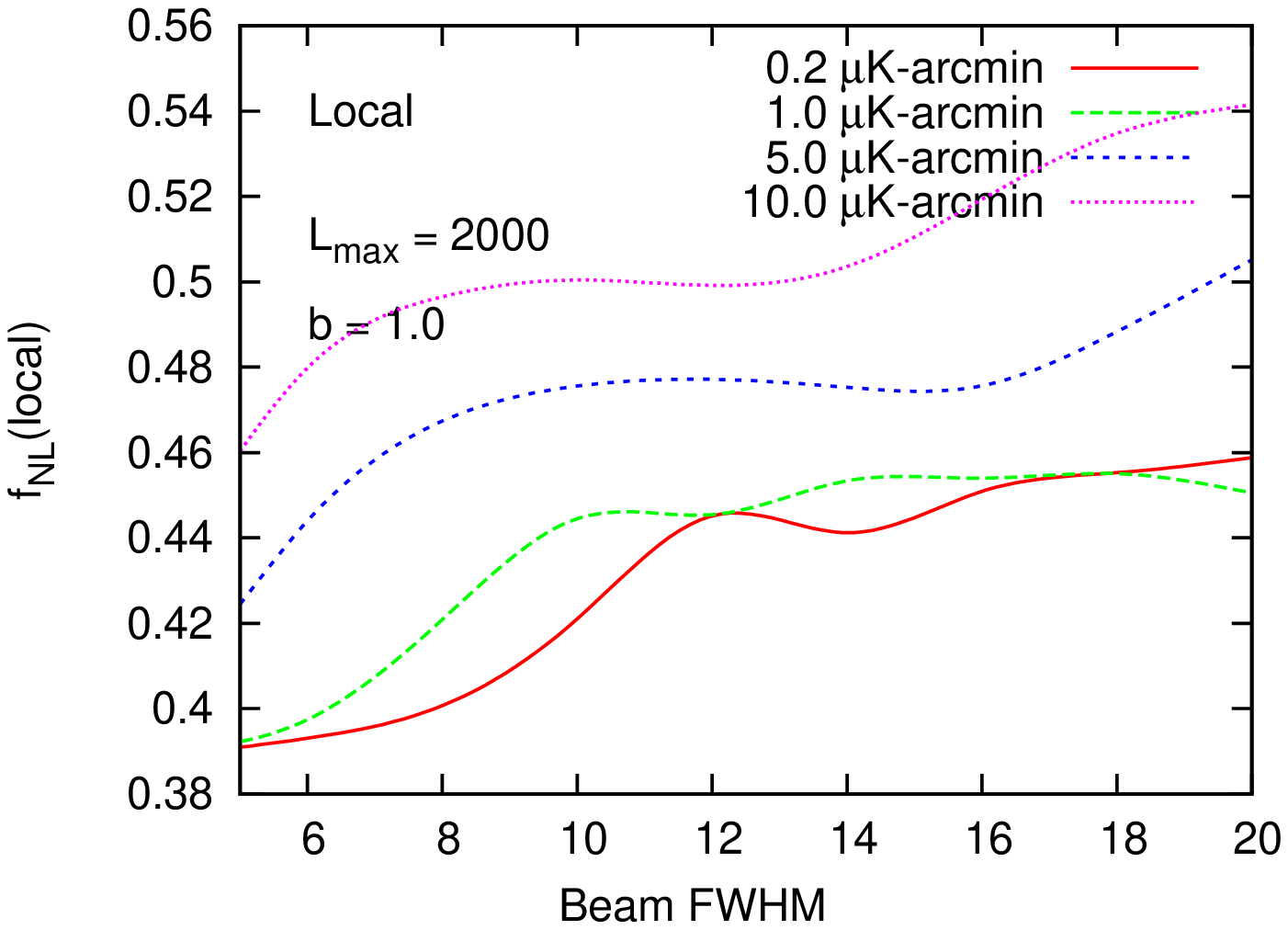}
\includegraphics[width=40mm,angle=-90]{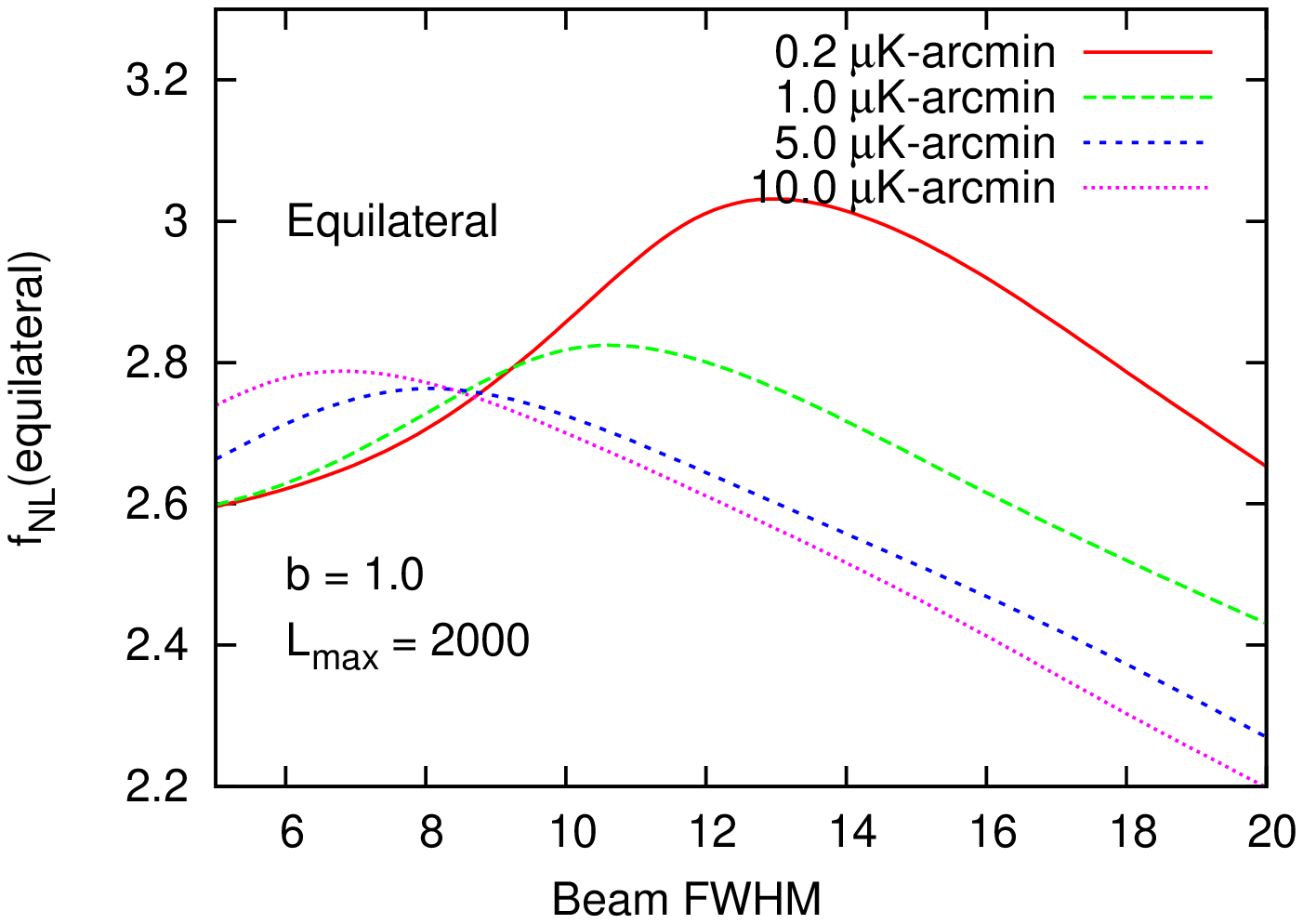}
\includegraphics[width=40mm,angle=-90]{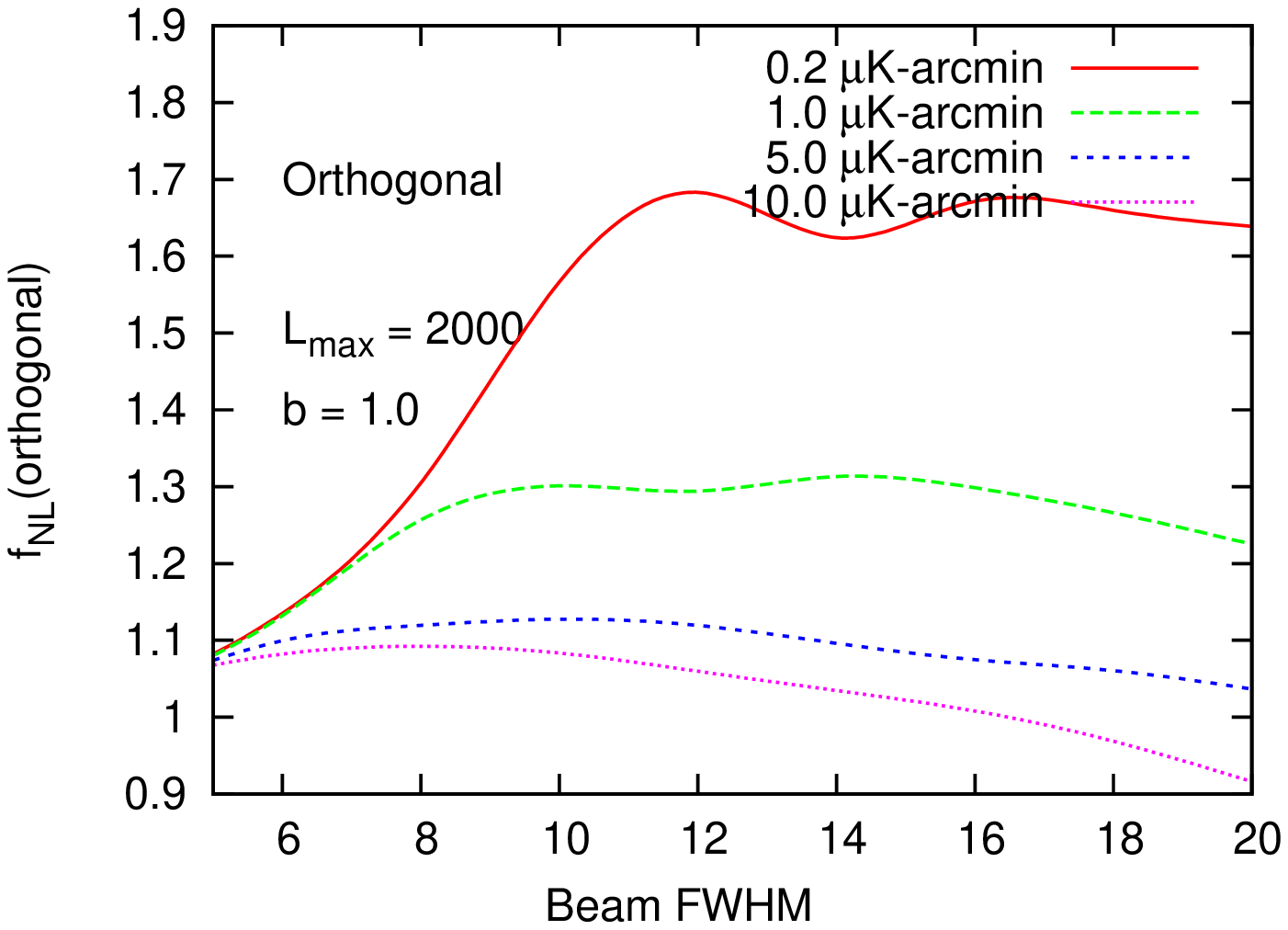}
\caption{{\it Upper panels:} The requirement on non-linear systematics parameter $b$ at $1\sigma$ level as a function of $FWHM$ for various choices of instrumental sensitivity and for local (left), equilateral (middle), and orthogonal (right) templates.  The requirement were derived by demanding that the spurious local $f^{syst.}_{\rm NL} $ generated by systematics are detectable at $1\sigma$ i.e.  by solving Eq.~(\ref{eqn:requirement}).
{\it Lower panels:} The minimum detectable effective $f^{syst.}_{\rm NL}$ from non-linear systematics for $b=1$ as a function of $FWHM$ and for 3 choices of instrumental sensitivity, for local (left), equilateral (middle), and orthogonal (right) templates.}
\label{fig:requirementT}
\end{figure} 


\section{Numerical Results}
\label{sec:results}

In order to numerically calculate the systematic effects on bispectrum, one needs to choose specific model of systematics field, i.e. the power spectrum $C^{SS}_{\ell}$ of the systematics field $S$. So far we have not made any assumption about the systematic field. In reality the power spectra of the systematics fields will depend on the experimental design and scan strategy, and could be inhomogeneous, anisotropic and complicated. However, we here employ a model which, although not exact, can be used to assess the level of contamination, independent of the origin of these systematics fields. As a simple model we assume that the contamination fields are statistically isotropic and Gaussian, thus their statistical properties can be fully described by their power spectra~\cite{HHZ},

\begin{equation}
\left< S (\bfl) S'(\bfl') \right> = (2\pi)^2 \delta(\bfl+\bfl') C_{\ell}^{SS'}\,,
\end{equation}
where $S$ stands for any of the systematic fields. We assume the power spectra of the form

\begin{equation}
C_{\ell}^{SS} = \frac{A_{S}^2 \exp(-l(l+1)\sigma_{S}^2)}{\int {d^2 l \over (2\pi)^2}
\exp(-l(l+1)\sigma_{S}^2)}\,, \label{eq:coh}
\end{equation}
i.e. white noise above certain coherence scale $\sigma_{S}$, which is a key quantity that sets the level of contamination of each systematics effect together with $A_S$, which describes the {\it rms} of the contamination field $S$. Note that $\sigma$ without the subscript $S$ refers to the instrumental beam $FWHM$.

In Fig~(\ref{fig1}) we show the systematics-induced temperature bispectrum $B^{TTT,S}_{(\bfl_1,\bfl_2,\bfl_3)}$ whose analytic result are given by Eq.~(\ref{eq:tempgeneration}) and~(\ref{eq:tempdistortion}). In the upper right and lower panels we show the effect of linear systematics $a$ which distort the CMB bispectrum of local, equilateral, and orthogonal templates. In each case we consider several choices of coherence length varying from $\sigma_s=10'$ to $\sigma_s=120'$. In the upper left panel we also show the systematic-induced bispectrum from the cross term $C_{\ell}^{ab}$ of Eq.~(\ref{eq:tempgeneration}). The systematics-induced bispectrum scales as $A^2_{s}$. In all the these panels the {\it rms} amplitude $(A_a=A_b=0.5)$ was set to high values for illustration purposes. For experiments with systematics {\it rms} fluctuations $A_S \sim 10\%$, the linear distortion is not the dominant effect for $\ell \lesssim 2000$ if coherence $\sigma_S > 10'$. 
For experiments which probe much smaller scales ($\ell \gtrsim 2000$), the extracted non-Gaussianity might be contaminated by linear systematics.
Note that for large coherence lengths both distortion bispectrum and bispectrum induced due to $C_{\ell}^{ab}$ have a similar shape as the primordial bispectrum. However, for small coherence length the convolution transfers power from large to small scales and the resulting distorted bispectrum is much flatter than the primordial bispectrum. For large coherence length the systematics power spectrum drops rapidly and does not have power over broad enough $\ell$ range to transfer power around. 

We want to quantify the maximum tolerable systematics (for a given experiment) below which they would cause no significant degradation of our ability to constrain primordial non-Gaussianity. We again focus on three templates, local, equilateral and orthogonal. The confusion of non-Gaussianity $\fnl$ from systematics with the primordial non-Gaussianity can be quantified by solving 
\begin{eqnarray}
\sum_{XYZ}\sum_{X'Y'Z'}\int {d^2 \ell_1 \over (2\pi)^2}  {d^2 \ell_2 \over (2\pi)^2}  {d^2 \ell_3 \over (2\pi)^2}\,\,B^{XYZ,syst}_{\ell_1 \ell_2 \ell_3}  \Big( \tilde C^{-1}_{\ell_1}\Big)^{XX'} \Big(\tilde C^{-1}_{\ell_2}\Big)^{YY'} \Big(\tilde C^{-1}_{\ell_3}\Big)^{ZZ'} B^{X'Y'Z',prim}_{\ellt}&=& \nonumber\\ &&  \hspace{-5cm} f^{syst}_{\rm NL} \sum_{XYZ}\sum_{X'Y'Z'}\int {d^2 \ell_1 \over (2\pi)^2}  {d^2 \ell_2 \over (2\pi)^2}  {d^2 \ell_3 \over (2\pi)^2}\,\, B^{XYZ,prim}_{\ell_1 \ell_2 \ell_3}  \Big( \tilde C^{-1}_{\ell_1}\Big)^{XX'} \Big(\tilde C^{-1}_{\ell_2}\Big)^{YY'} \Big(\tilde C^{-1}_{\ell_3}\Big)^{ZZ'} B^{X'Y'Z',prim}_{\ellt},
\end{eqnarray}
where $\tilde C^{XX'}_{\ell}$ is the observed CMB power spectrum with noise. Here $f^{syst}_{\rm NL}$ is an effective amplitude of non-Gaussianity generated for a given template shape. We calculate the requirement on non-linear systematics errors by demanding that the effective $f^{syst}_{\rm NL}$ is not detectable at $1\sigma$ using the optimal CMB estimator as developed in Ref.~\cite{KSW05,creminelli_wmap1,Yadav_etal08a,Smith:2009jr}, i.e. $f^{syst}_{\rm NL} <\langle \fnl^2 \rangle^{1/2}$, giving the requirement
\begin{eqnarray}
b^{shape} < \frac{\sqrt{\int \frac{d^2l_1}{(2\pi)^2}  \frac{d^2l_2}{(2\pi)^2}   \frac{d^2l_3}{(2\pi)^2}  \frac{\Big( B^{shape,prim}_{\ell_1 \ell_2 \ell_3}\Big)^2}{\tilde C_{\ell_1}\tilde C_{\ell_2}\tilde C_{\ell_3}}}} { \int \frac{d^2l_1}{(2\pi)^2}  \frac{d^2l_2}{(2\pi)^2}   \frac{d^2l_3}{(2\pi)^2}  \frac{B^{bias}_{\ell_1 \ell_2 \ell_3} B^{prim, shape}_{\ell_1 \ell_2 \ell_3}}{\tilde C_{\ell_1}\tilde C_{\ell_2}\tilde C_{\ell_3}}}\,.
\label{eqn:requirement}
\end{eqnarray}
In Fig.~(\ref{fig:requirementT}) upper panels we show the maximum allowed non-linear temperature systematics $b$ as a function of beam $FWHM$ for local, equilateral and orthogonal shapes and for various choices of experimental noise sensitivity. For a cosmic-variance-limited experiment up-to $\ell_{max}=2000$, the requirement for local, equilateral and orthogonal shapes are $b<22\,\, (2\sigma)$, $b<41\,\, (2\sigma)$ and $b<44\,\, (2\sigma)$ respectively. We will see in the next section that these requirements from bispectrum are more stringent than the requirements for non-linear systematics from the $B$-mode power spectrum.

\begin{figure*}[t]
\includegraphics[width=41mm,angle=-90]{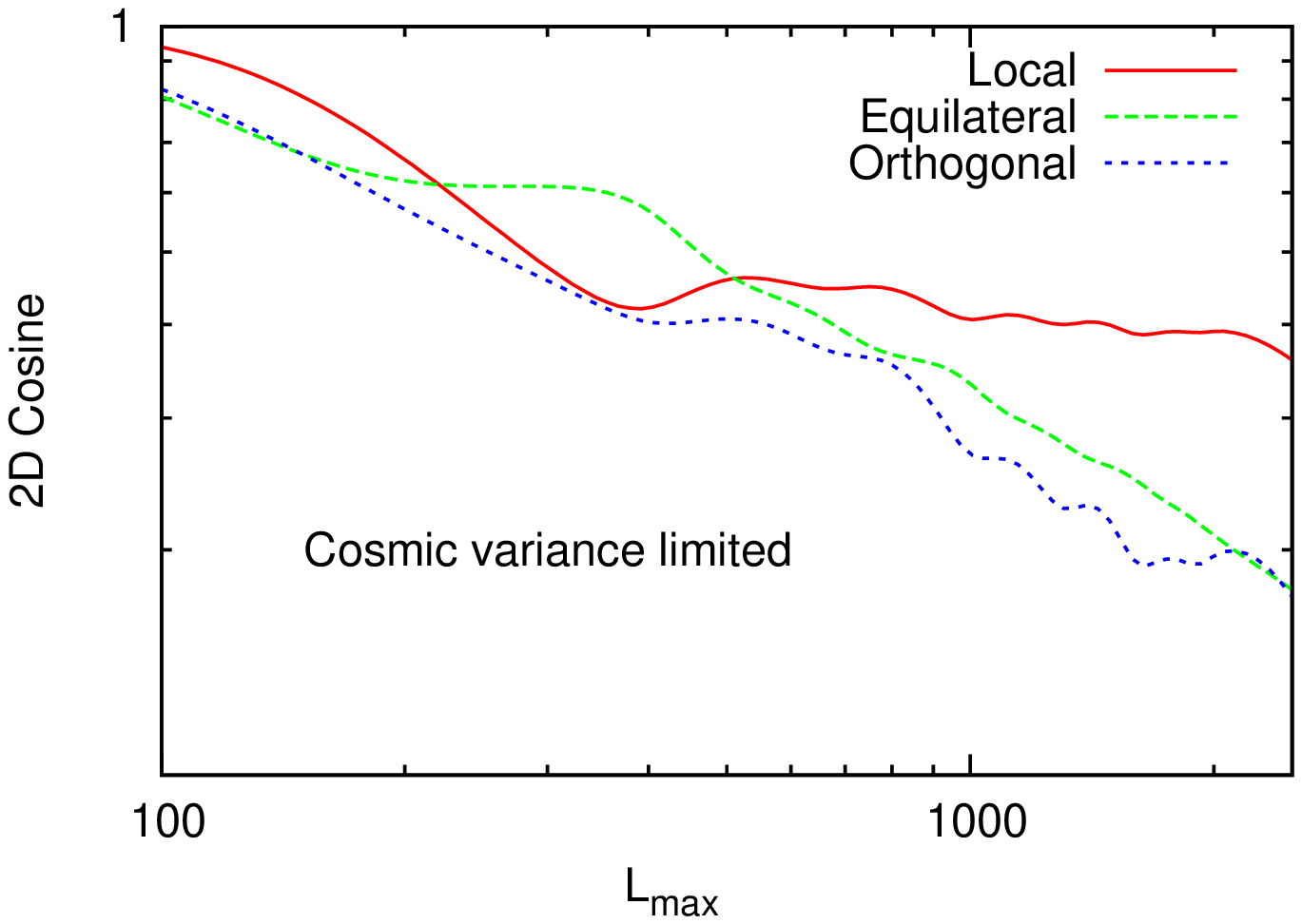}
\includegraphics[width=41mm,angle=-90]{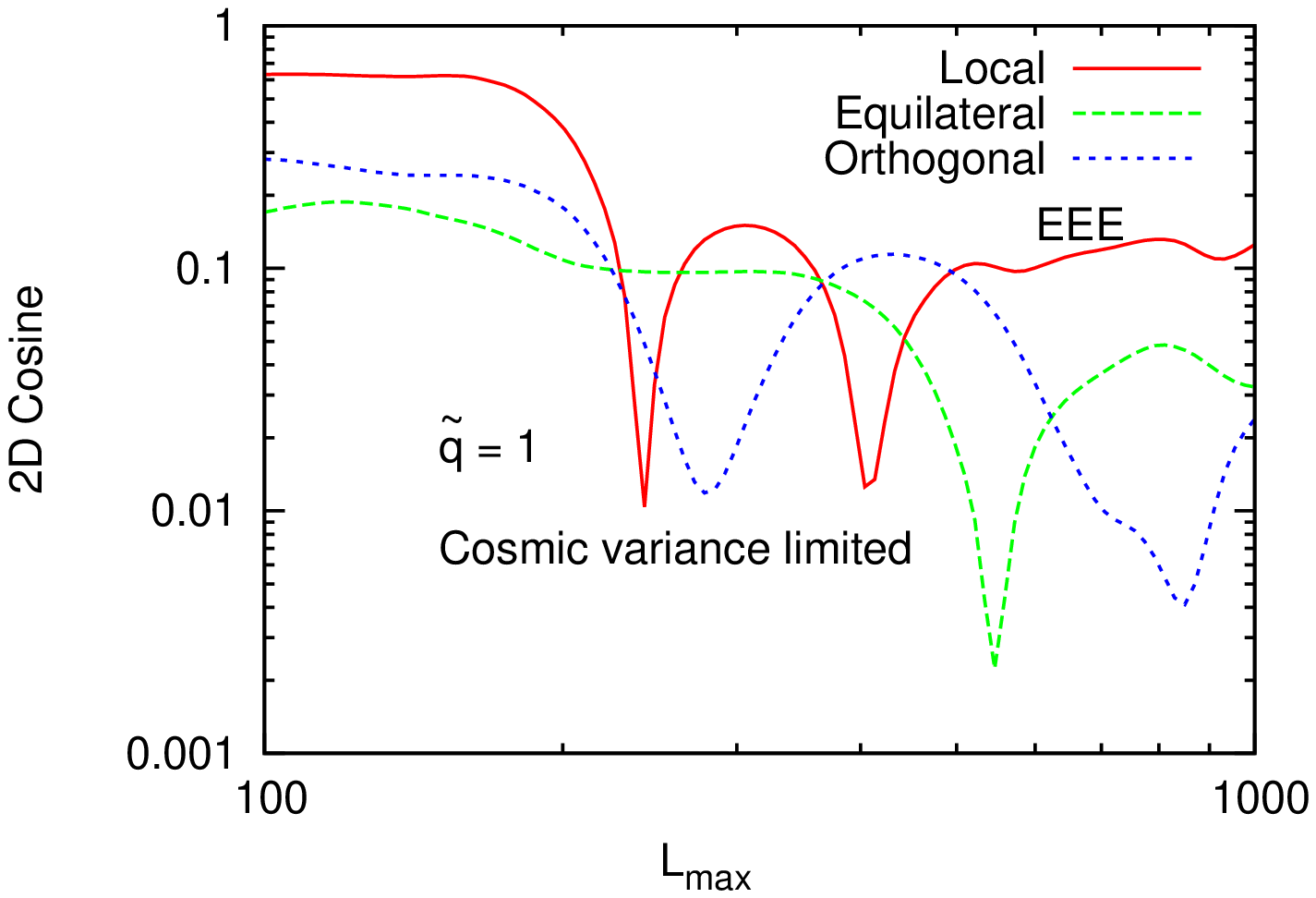}
\includegraphics[width=41mm,angle=-90]{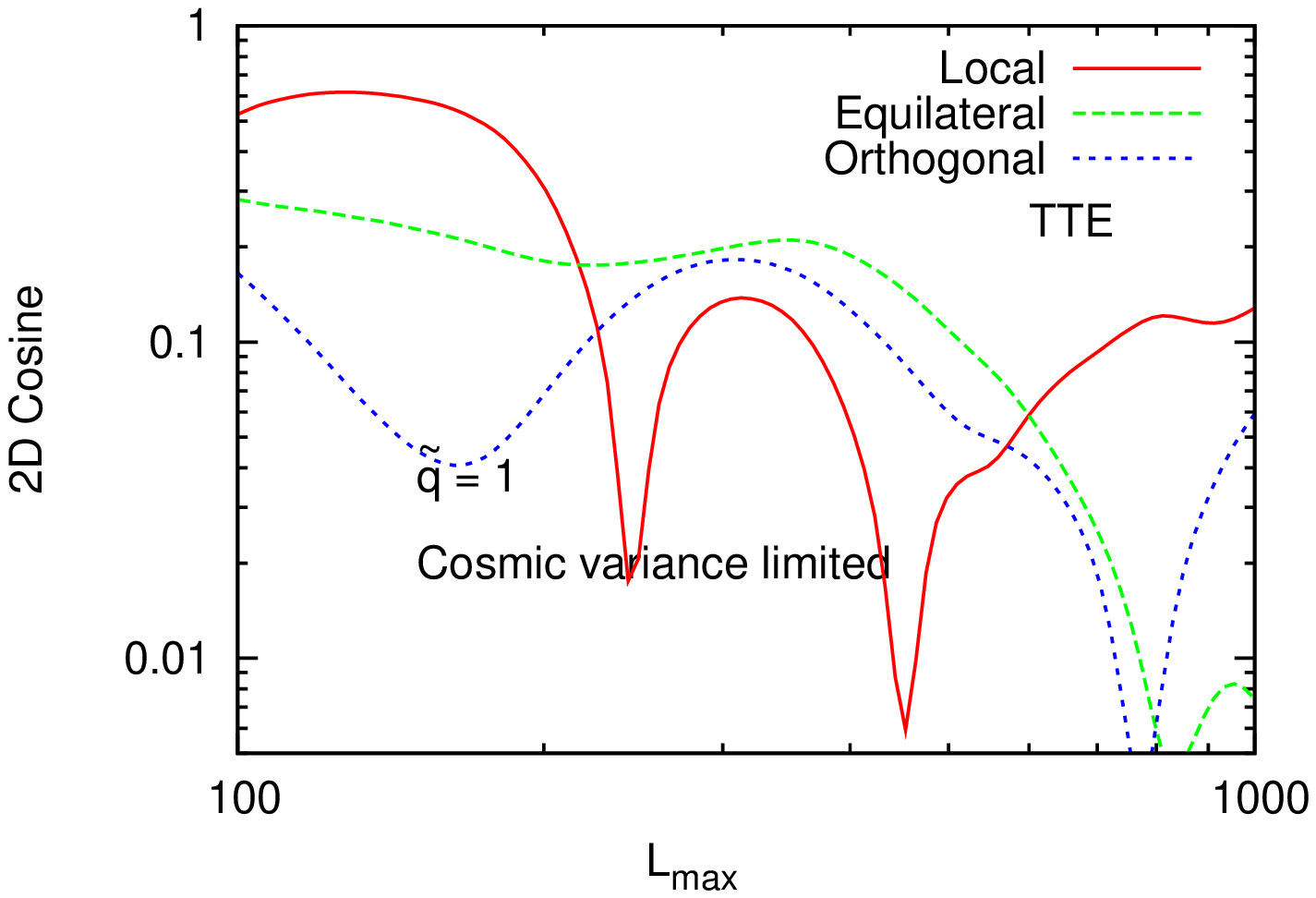}
\caption{2-d cosine between the non-linear systematics induced bispectrum and local, equilateral, and orthogonal template. The {\it left panel} is for the temperature systematics bispectrum $B^{TTT,b}$ and the {\it middle} and {\it right panels} are for the polarization $B^{EEE,\tilde q}$, and $B^{TTE,\tilde q}$, respectively. For the polarization plots (middle and right), the spikes in the plots correspond to sign-flip in the cross-correlation of systematics and primordial bispectrum. }
\label{fig:2dcosineT}
\end{figure*}

In Fig.~(\ref{fig:requirementT}) lower panels show the effective $f^{syst.}_{\rm NL}$ from the non-linear systematics for local, equilateral and orthogonal templates as a function of beam $FWHM$ and for several choices of experimental noise sensitivity. Note that systematics-induced $f^{syst.}_{\rm NL}$ is only weakly sensitive to the instrument beam and detector sensitivity. Also since $f^{syst.}_{\rm NL}$ is of order unity for $b=1$, the requirement on $b$ is comparable to minimum detectable $\fnl $ i.e. $\fnl \sim b$ for given CMB experiments. As expected the requirement on the systematic control gets weaker as the $FWHM$ and/or instrumental noise is increased. Also note that since the maximum multipole used in the analysis was fixed to $\ell_{max}=2000$, all experiments with low enough instrument noise and $FWHM$ give identical requirement because they all are cosmic variance limited up-to $\ell_{max}=2000$, as indicated in Fig.~(\ref{fig:requirementT}).

To quantify the correlation between the systematics-induced bispectrum and primordial bispectrum template, one can calculate the 2-d cosine as defined in Eq.~(\ref{eqn:2dcosine}). Fig.~(\ref{fig:2dcosineT}) shows this 2-d cosine, between the systematics and local, equilateral and orthogonal templates. Systematics have comparable overlap with all the three templates, and overlap is relatively small with 2-d cosine of order $\sim O(0.2)$ for equilateral and orthogonal shape at $\ell_{max}\sim 2000$, and of order $\sim O(0.4)$ for local shape. Since we modeled the temperature non-linear response as $\cmb(\bn)=\cmb(\bn)+b\cmb(\bn)^2$, one might expect a large overlap with local template however this not the case because (1) the local template is defined for three dimensional $\bfk$ space of primordial curvature perturbations, while we show the overlap with the projected two dimensional $\ell-$space bispectrum;  (2) at high $\ell$ there are oscillatory features in both the template and systematics bispectrum which after summation over $\ell$ cancels out. Because of the week overlap between systematics and primordial templates, systematics generate comparable amount of non-Gaussianity for the three templates considered.

Given the systematics contamination $f^{syst.}_{\rm NL}$ for a given shape, one can get a rough estimate for the level of $f^{syst.}_{\rm NL}$  for another shape by using the 2d-cosine between the two shapes. For example, if we start from systematics contamination for equilateral shape for a cosmic variance limited experiment, $f^{syst,equil}_{\rm NL} \approx 2.6$ (see Fig.~(2) lower middle panel). We can convert it to the effective local $f^{syst,local}_{\rm NL}$ by using a fudge factor $\sim 6$ between the two shapes to get $f^{syst,local}_{\rm NL} \approx 2.6/6=0.44$. This agrees nicely (see Fig. 2 lower left panel)  with what we get with the full calculation. The fudge factor $6$ comes from 2D cosine which gives a factor of $2$ and the normalization which gives a factor $3$~\cite{Babich_etal_04}. The reason we get higher $\fnl$ in the equilateral model is that we normalize the maximum of this shape to the minimum of the local shape. The normalization factor 3 compensates for this effect\footnote{The local and equilateral shapes are normalized to the same value at equilateral configuration, i.e. $F_{local}(\bfk,\bfk,\bfk)=F_{equil}(\bfk,\bfk,\bfk)$} (see Ref.~\cite{Babich_etal_04} for details).


For polarization, many of the instrumental systematics can be removed or suppressed by carefully designing the scan strategy. Instrumental rotation is potentially a powerful way to mitigate instrumental systematics effects if the systematics field has different spin-dependence than the CMB polarization $Q\pm iU$. Ideally, if every pixel can be covered by infinite number of different rotation angles with respect to the real polarization direction on the sky, systematics fields ($\gamma_1, \gamma_2, d_1, d_2, \omega, f_1, f_2, p_1,$ and $p_2$) which have different spin from $Q\pm iU$ can be completely removed. However the quadrupole leakage (leakage from temperature to polarization, stemming from beam ellipticity) has the same spin dependence as the polarization parameters $Q\pm iU$ and hence can not be averaged over, even with the perfect coverage of rotation angles. For this reason we mainly focus of quadrupole leakage $\tilde q$. As with the temperature systematics $b$, we find that only the monopole of systematics field can generate new bispectrum. Fluctuations can only distort the existing bispectrum without generating spurious signals.

\begin{figure}[t]
\includegraphics[width=41mm,angle=-90]{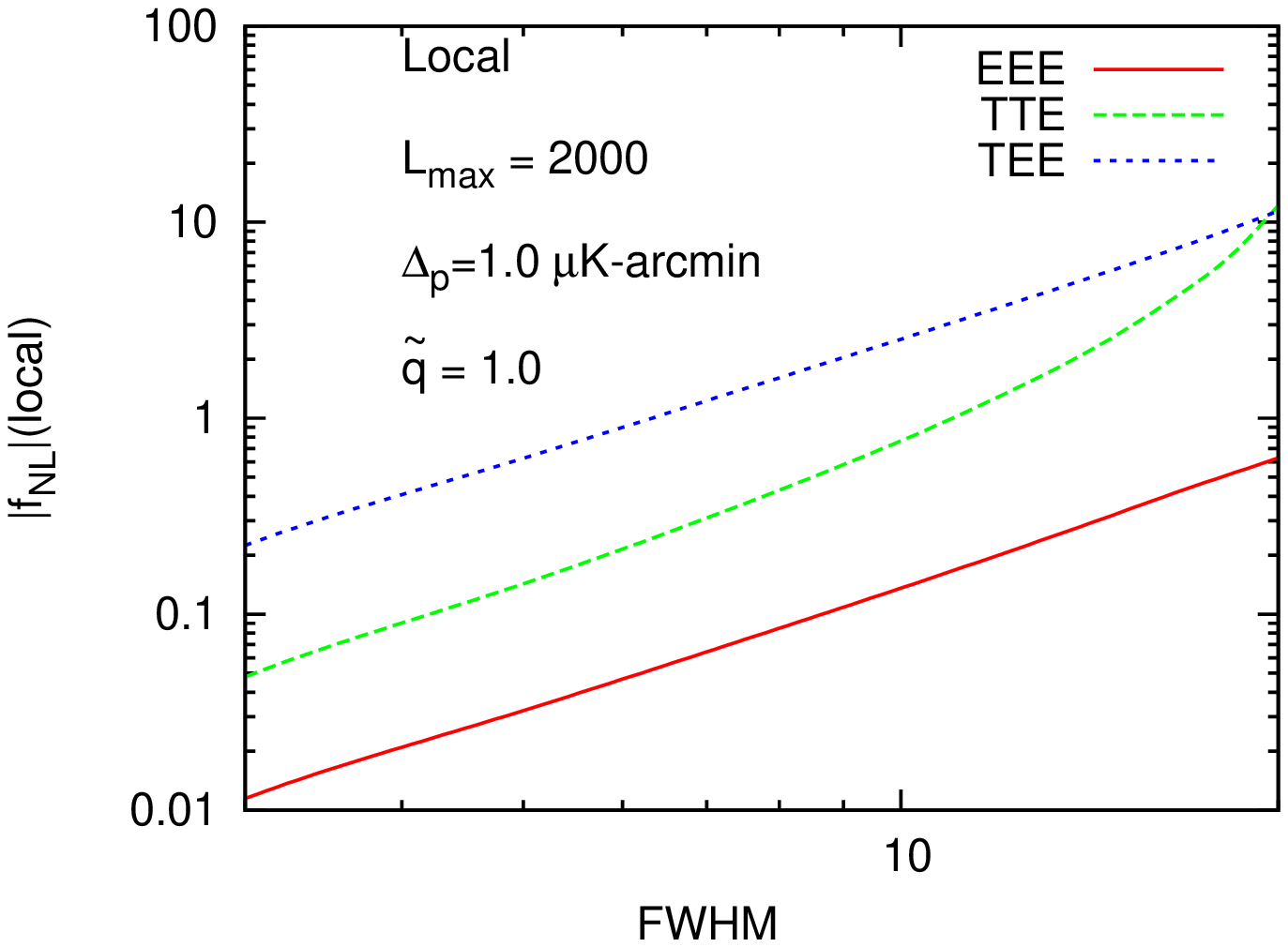}
\includegraphics[width=41mm,angle=-90]{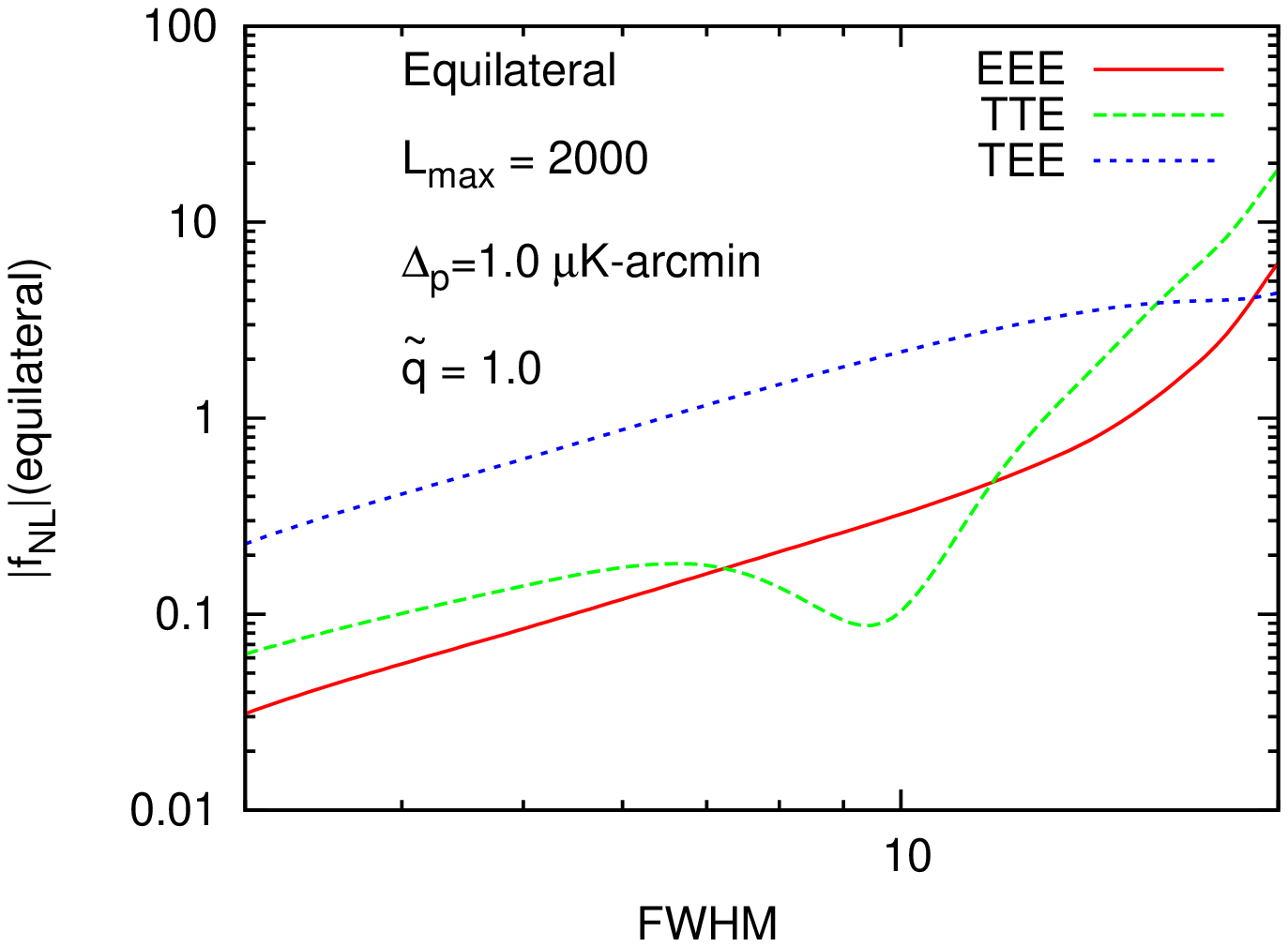}
\includegraphics[width=41mm,angle=-90]{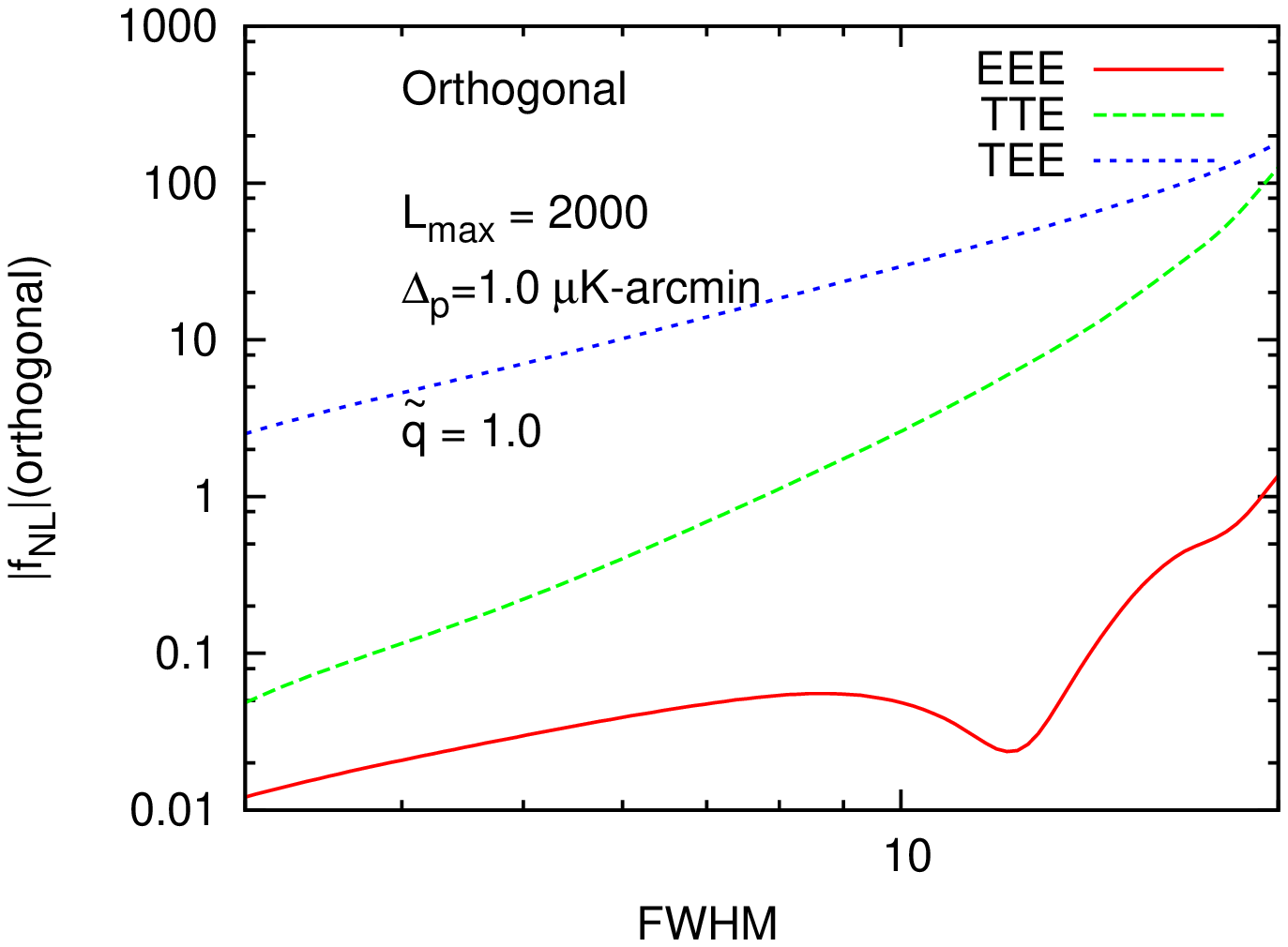}
\includegraphics[width=41mm,angle=-90]{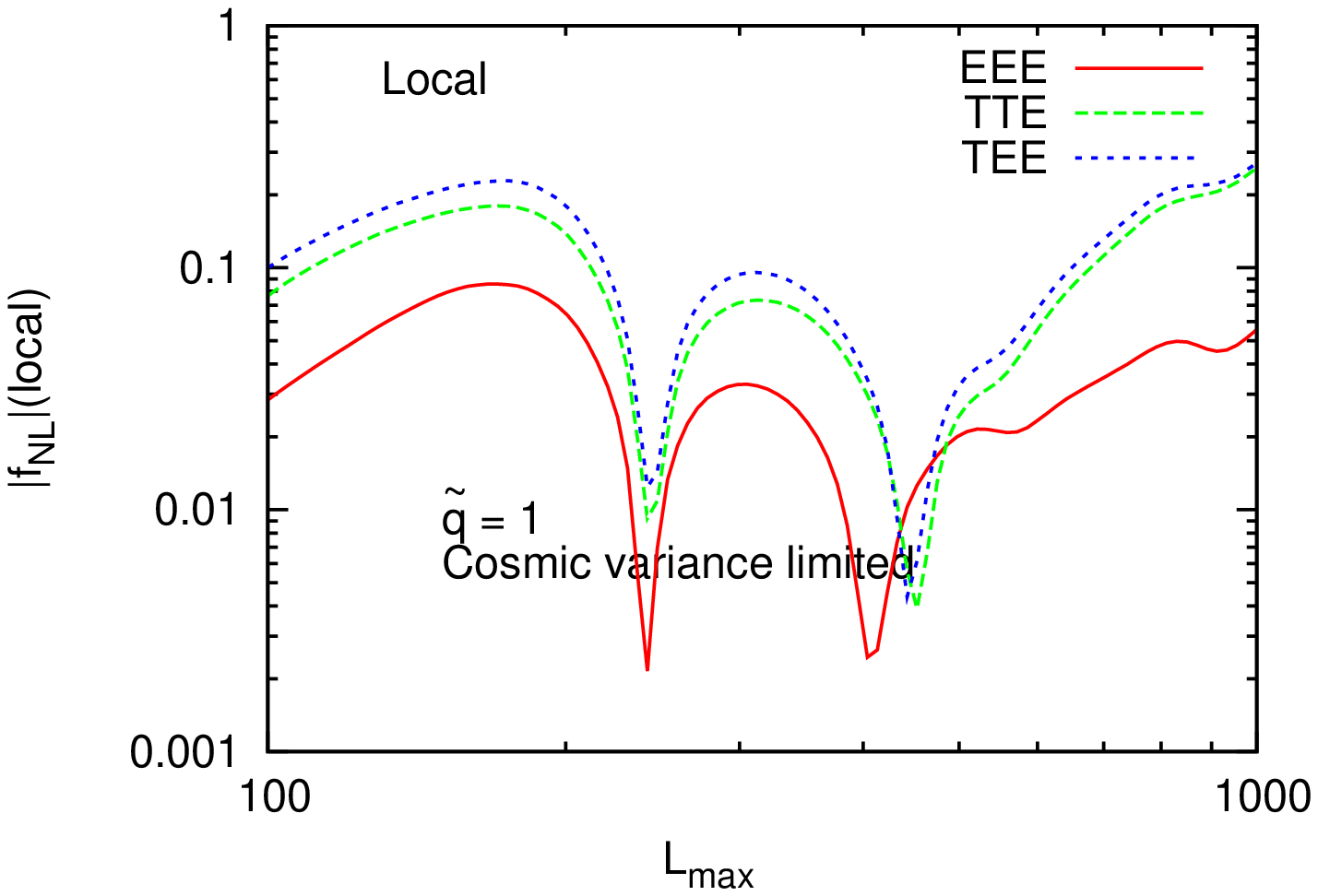}
\includegraphics[width=41mm,angle=-90]{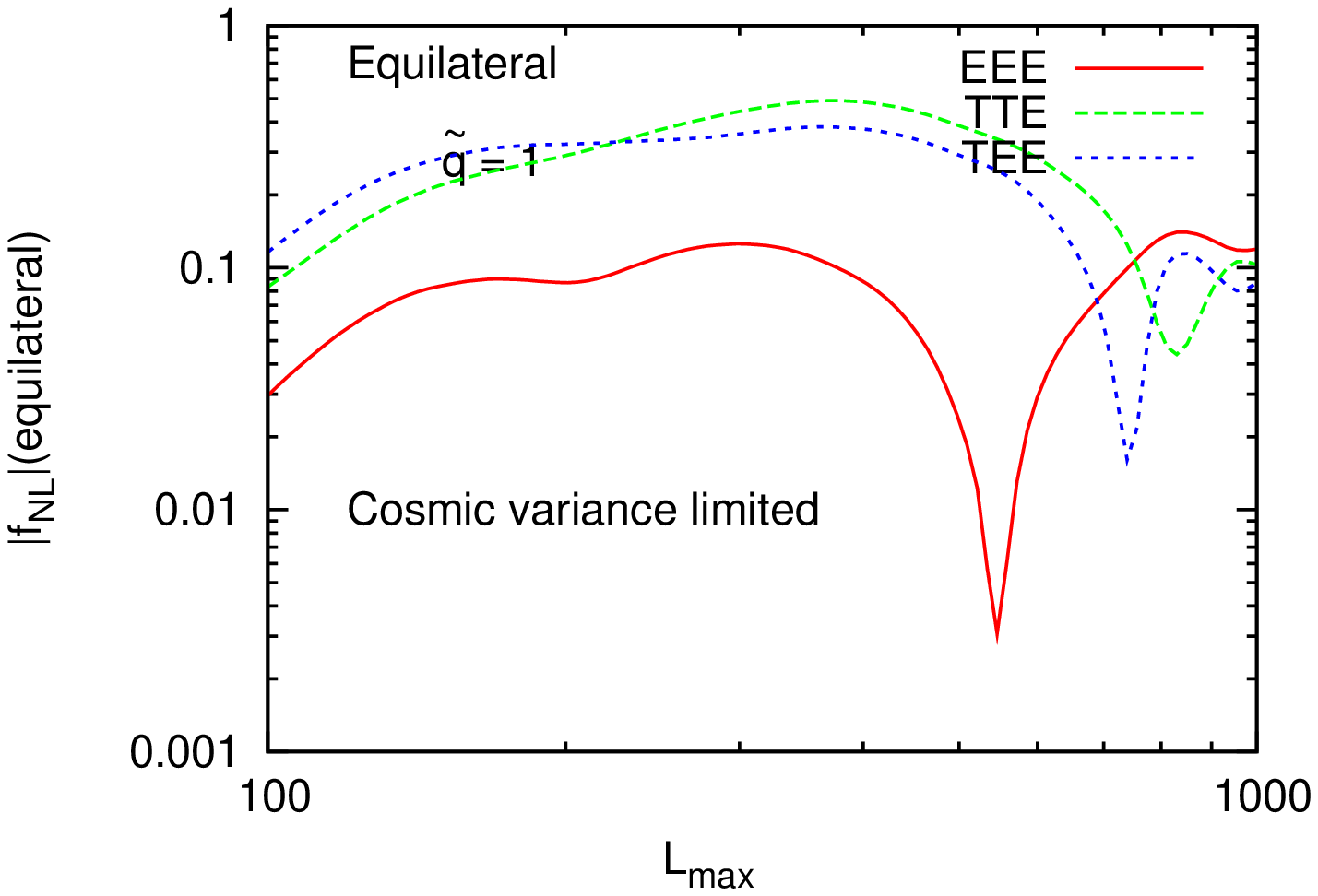}
\includegraphics[width=41mm,angle=-90]{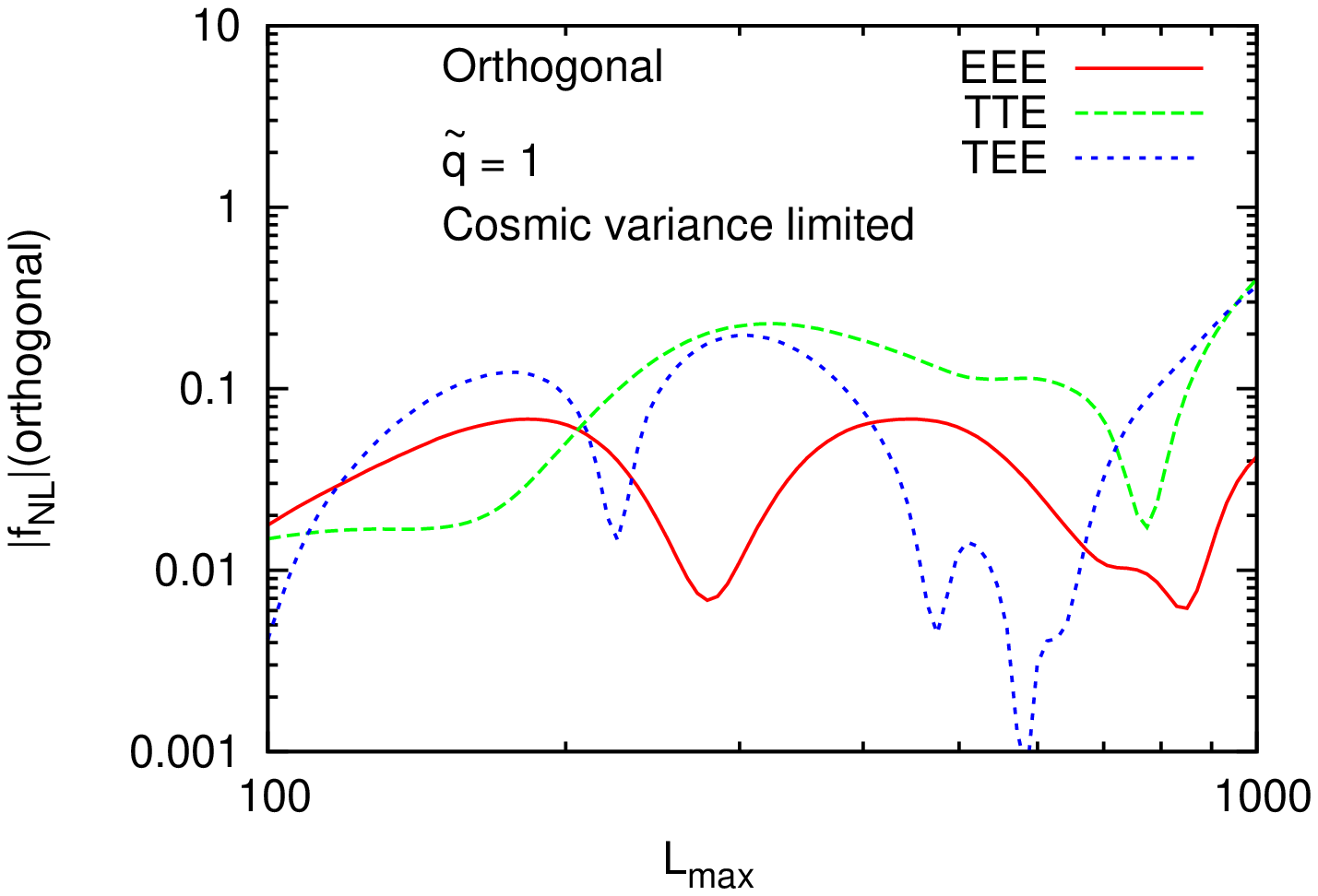}
\caption{Effective $f_{\rm NL}$ of local (left), equilateral (middle), and orthogonal (right) templates generated from non-linear quadrupole leakage $\tilde q$ as function of $FWHM$ ({\it upper panels}) and $\ell_{max}$ ({\it lower panels}) for EEE, TEE,TTE estimators. For {\it lower panels}  we assume a cosmic-variance-limited experiment. The spikes in the plots correspond to sign-flip in the cross-correlation of systematics and primordial bispectrum.}
\label{fig:Pol1}
\end{figure}

In Fig~(\ref{fig:Pol1}) we show the effective non-linearity parameter $f^{syst.}_{NL}$ generated by quadrupole leakage $\tilde q$ for the local, equilateral and orthogonal templates. We show the  $f^{syst.}_{NL}$ as function of $FWHM$ (upper panels) and as a function of maximum multipole used in the analysis $\ell_{max}$. Since the quadrupole leakage come from the beam effects, larger the beam $FWHM$, larger the spurious non-Gaussianity $f^{syst.}_{NL}$. The spikes in the $\fnl$ vs $\ell_{max}$ plots correspond to sign-flip in the cross-correlation of systematics and primordial bispectrum.

\begin{figure}[t]
\includegraphics[width=41mm,angle=-90]{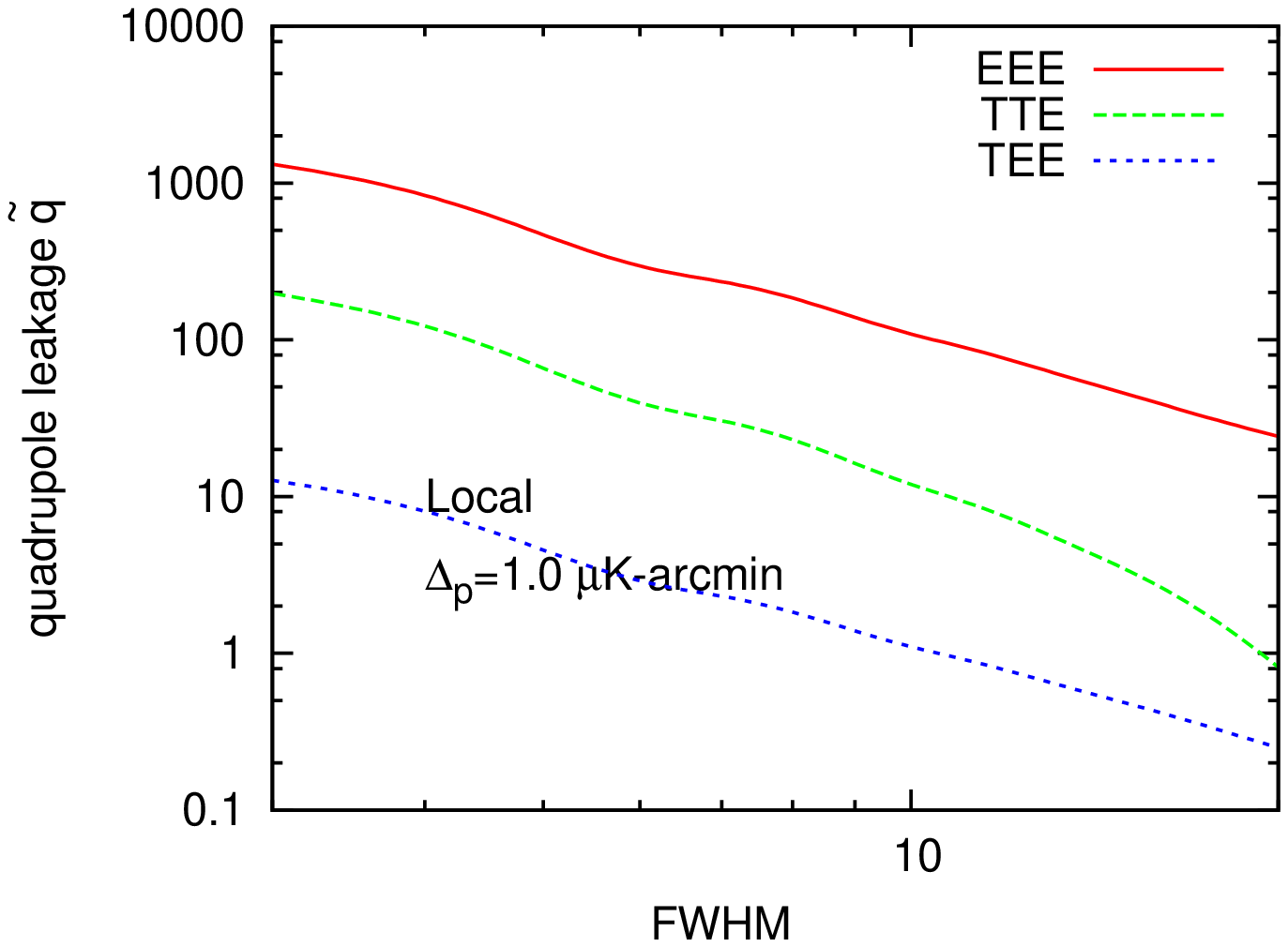}
\includegraphics[width=41mm,angle=-90]{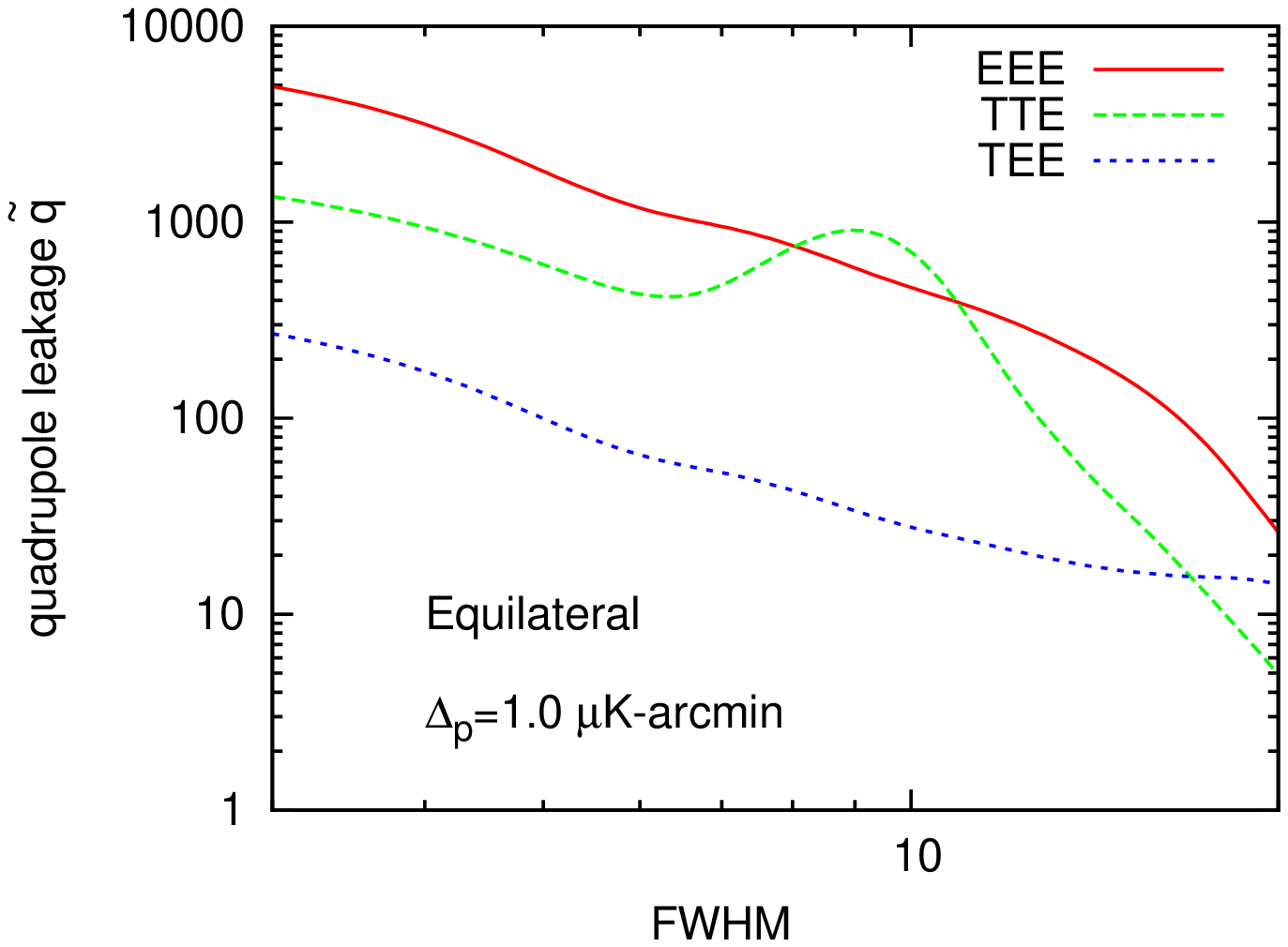}
\includegraphics[width=41mm,angle=-90]{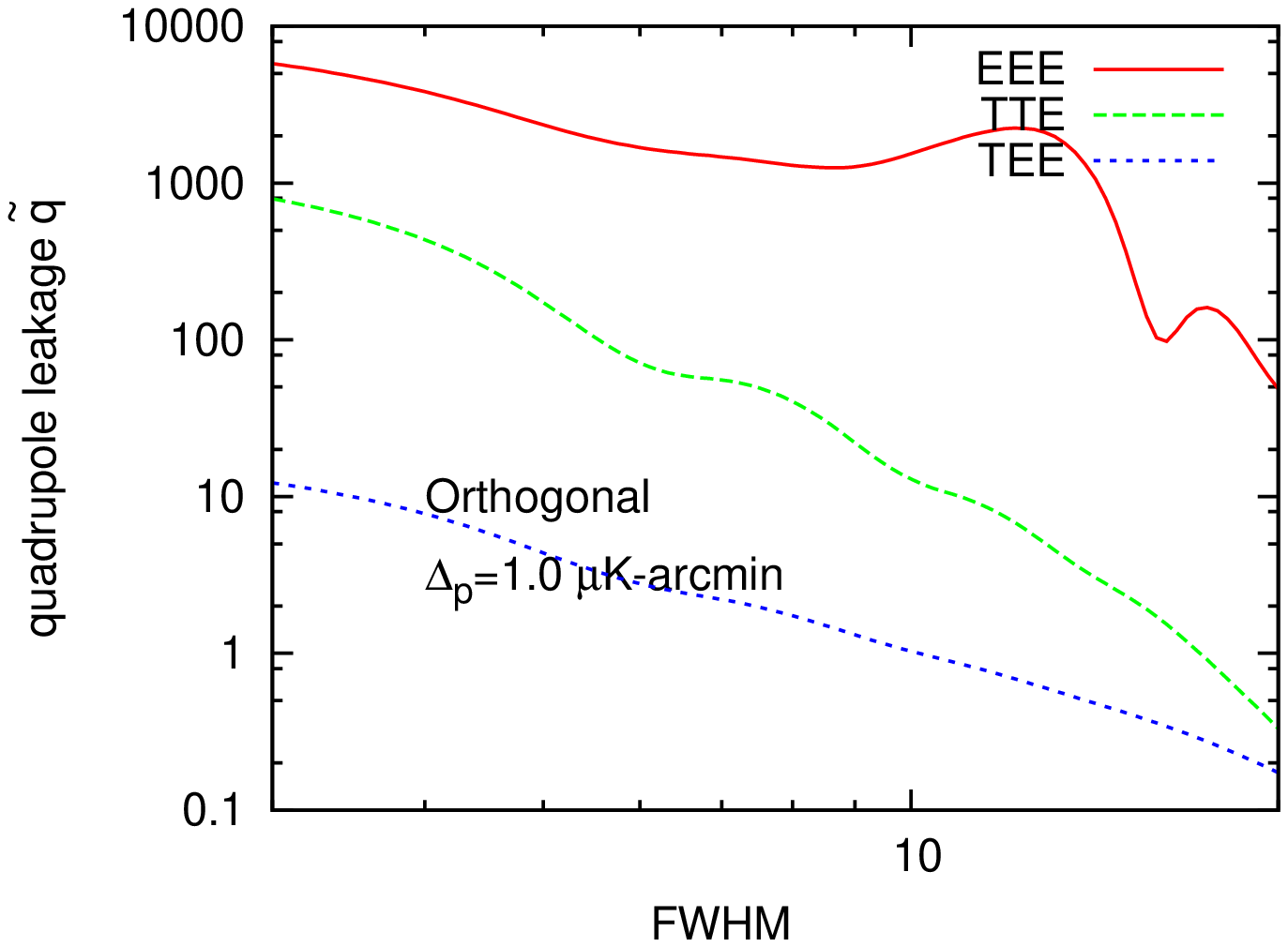}
\includegraphics[width=41mm,angle=-90]{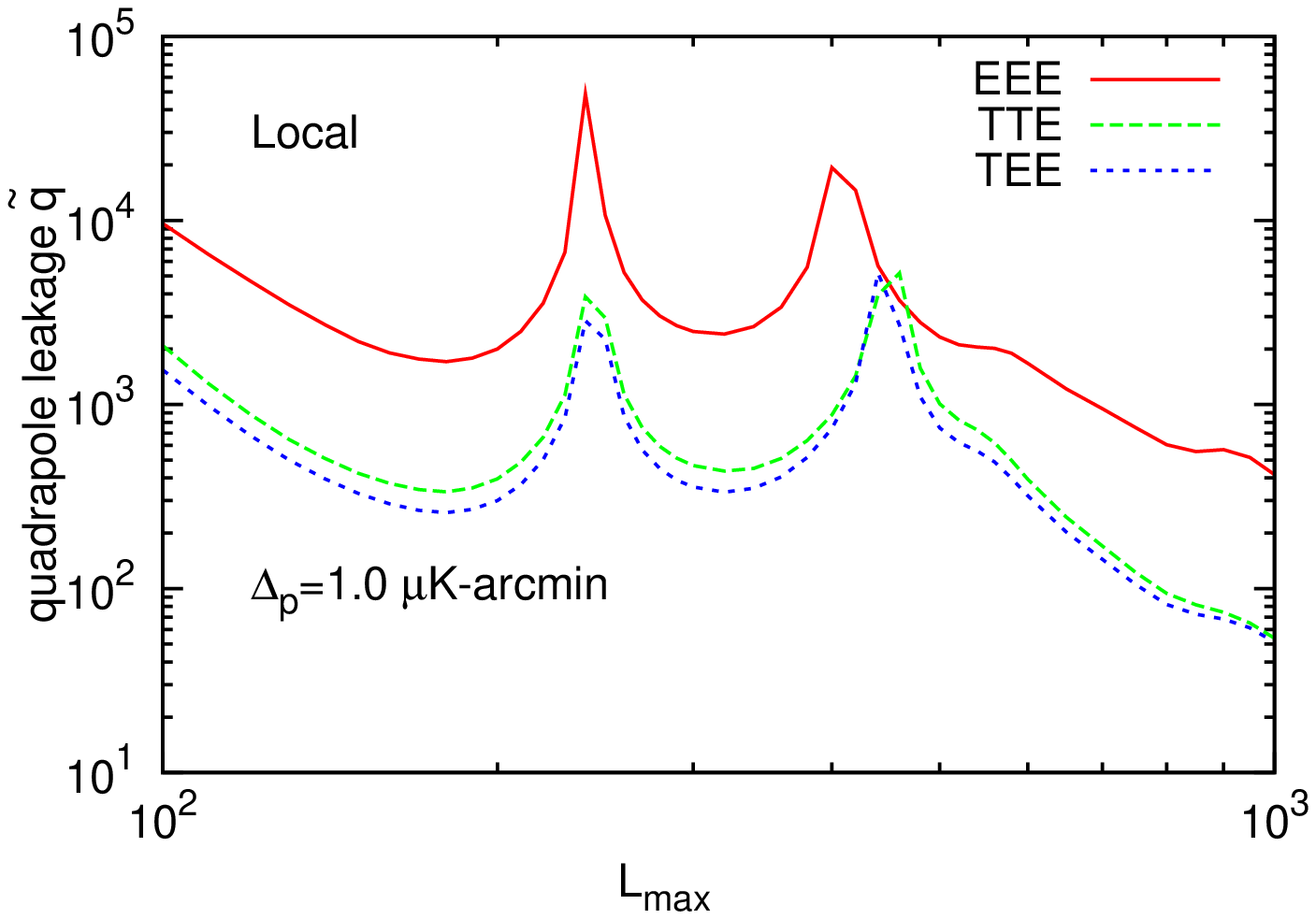}
\includegraphics[width=41mm,angle=-90]{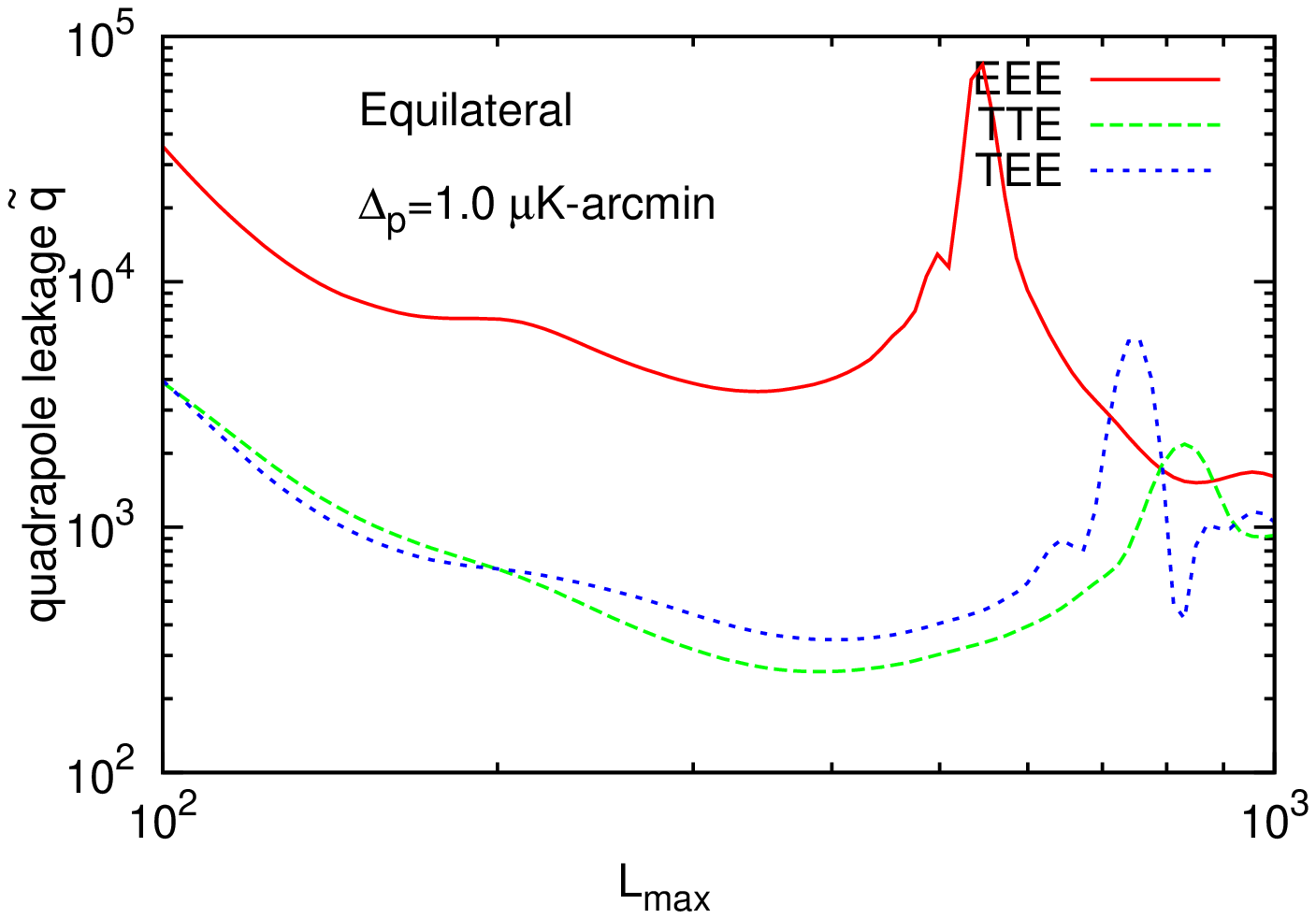}
\includegraphics[width=41mm,angle=-90]{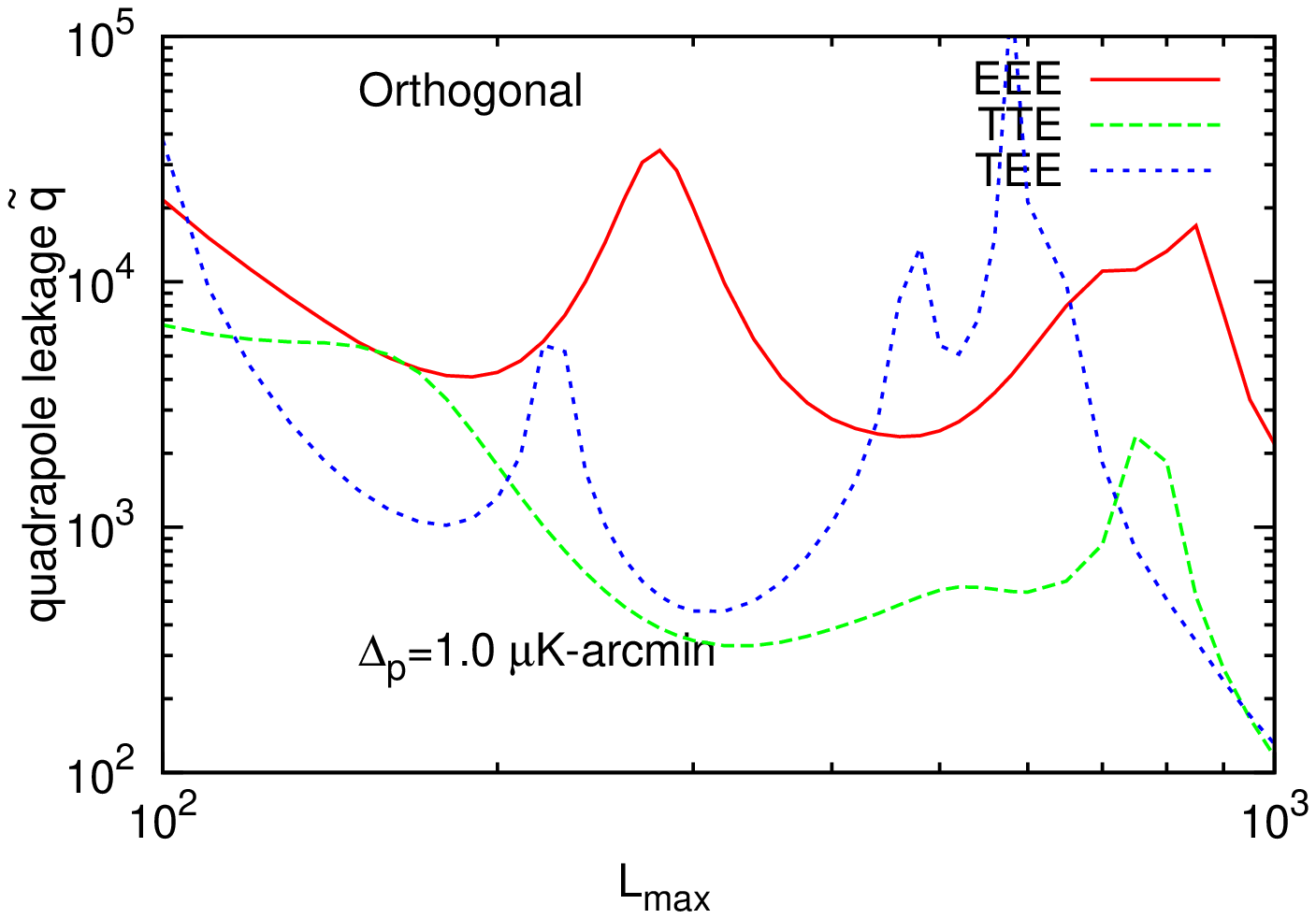}
\caption{Requirements from the $TTE, TEE$ and $EEE$ bispectrum for the non-linear quadrupole leakage $\tilde q$ as function of $FWHM$ (upper panels) and $\ell_{max}$ (lower panels) for noise sensitivity $\Delta_p=1\mu$K-arcmin and for local (left), equilateral (middle) and orthogonal templates (right). The requirements were derived by demanding that the spurious local $\fnl$ generated by systematics are detectable at $1\sigma$. Note that from the definition of quadrupole leakage, the requirements are expressed in the units of $\sigma^2$ and therefore the requirement becomes more stringent as the beam $FWHM$ is increased. For lower panels the beam $FWHM$ is $10'$, and the spikes correspond to sign-flip in the cross-correlation of systematics and primordial bispectrum.}
\label{fig:Pol2}
\end{figure}

In (\ref{fig:Pol2}) we show requirements on non-linear polarization systematics parameter $\tilde q$ using separately the $TTE, TEE$ and $TEE$ bispectrum. As for the temperature case, the requirements were derived by demanding that the spurious local $f^{syst.}_{NL}$ generated by systematics are detectable at $1\sigma$. Note that from the definition of systematics leakage the requirements for quadrupole leakage $\tilde q$ are expressed in the units of square of beam $FWHM$.  In comparison to the temperature systematics the requirement from polarization is much weaker. Since the polarization systematics bispectrum contains sines and cosines, the dot product between the template bispectrum and polarization bispectrum oscillates, resulting in weaker effective $f^{syst.}_{NL}$. Since the quadrupole leakage scales as $\tilde q \sim \sigma^2$, the effective $f^{syst}_{NL}$ shown in Fig~(\ref{fig:Pol1}) also has the same scaling.

\section{Implications for CMB experiments}
\label{sec:Bmode}
How well can non-linear systematics be controlled? The systematics requirements on the linear response parameter $a$ for CMB experiments aimed at $B$-mode observations is well discussed in the literature~\cite{HHZ,SKPH}
\begin{eqnarray}
a &<& 0.06 \,\Big(\frac{r}{0.005}\Big)^{1/2}\, \text {for coherence}\,\, \sigma_a=10' \nonumber \\
a &<& 0.05 \, \Big( \frac{r}{0.005}\Big)^{1/2}\, \text{for coherence}\,\, \sigma_a =120'
\label{eq:scale}
\end{eqnarray}
To estimate the level of non-linear response $b$ which can be detected at the power-spectrum level, we start with  
\begin{equation}
\delta \cmb^{obs}(\bn)= (1+a) \delta \cmb^{sky}(\bn) + b \Big(\delta \cmb^{sky}(\bn)\Big)^2\,,
\end{equation}
and note that the non-linear term is negligible for the limits when $\delta \cmb^{sky}\rightarrow 0$ and contributes the most when $\delta \cmb^{sky}= \delta \cmb_{max}$. Therefore, given the maximum allowed linear response parameter $a$, one can obtain a requirement on $b$ as
\begin{equation}
b \le \lambda \frac{a_{max}}{\delta \cmb_{max}}
\end{equation}
where $\lambda$ is a fudge factor of order unity. Primary CMB anisotropy will provide an upper bound of $b \approx a\times 10^{5}$ from Eq.~(\ref{eq:scale})~\cite{HHZ}. The requirements on $a$ are more stringent from $B$-modes than from bispectrum systematics. Using the numbers we get 
\begin{eqnarray}
b &<& 6000 \Big(\frac{r}{0.005}\Big)^{1/2}\, \text {for coherence}\,\, \sigma_a=10' \nonumber \\
b &<& 5000 \Big( \frac{r}{0.005}\Big)^{1/2}\, \text{for coherence}\,\, \sigma_a =120'\,.
\label{eqn:b_requirement_powerspectra}
\end{eqnarray}
However, we can go beyond this rough estimate; one can use the dipole of the CMB to set more stringent requirement on $b$. With $\cmb_{dipole}\sim 0.001$ one obtains a requirement which is $\sim 100$ times more stringent than provided by Eq.~(\ref{eqn:b_requirement_powerspectra}). For $r=0.005$, this requires $b<50$, which is still less stringent than the requirement from bispectrum. Therefore, if instrumental non-linearities are not controlled, the effective non-Gaussianity can be as large as $\fnl\sim 30$ (see figure~2) before the corresponding non-linearities show up in CMB dipole. The higher multipoles of the CMB power-spectrum are even less sensitive to experimental non-linearities.

\subsubsection*{Non-linear response of the PLANCK satellite}
\label{sec:Planck}
PLANCK is the third generation space radiometer, following the COBE-DMR and WMAP instruments, and is currently collecting data from the Lagrangian point L2. PLANCK is expected to significantly improve our knowledge on non-Gaussianity of the CMB.  Here we discuss the the non-linear response of the Planck detectors, following Ref.~\cite{Mennella2009,Mandolesi2010,Leahy2010}. Two state-of-the-art and complementary instruments are integrated in the focal plane of the PLANCK 1.5-meter dual reflector telescope. The Low Frequency Instrument (LFI), based on coherent receivers, observes the sky in three bands centered at 30, 44 and 70 GHz. The HFI, using bolometers cooled down to 0.1 K, covers six channels between 100 and 850 GHz. The LFI is an array of 22 cryogenic coherent differential radiometers based on indium phosphide HEMT (high electron mobility transistor) low noise amplifiers. The receiver array is split into a front-end unit (cooled down to 20 K to optimize sensitivity) and a back-end unit (operating at $\sim$300 K). The instrumental properties of the LFI and HFI have been calibrated and tested at different integration levels by detectors, individual receivers, and the whole receiver array during ground calibration and in flight calibration. For LFI, the response of 30 and 40 GHz radiometers show slight output compression, which affects the noise properties. The effect on noise temperature, white noise sensitivity, and noise effective bandwidth has been studied extensively. Here we focus on the implications on bispectrum measurement. 

The non-linearity of microwave receiver can be introduced by various sources, e.g. radio-frequency (RF) amplifiers, detector diode, and back-end analog electronics~\cite{Mennella2009}. The back-end RF amplifiers and detector diodes might cause the non-linear response of 30 and 40 GHz receivers on LFI. For the LFI pseudo-correlation receivers, assuming the sky signal and the reference load can be perfectly isolated after the hybration
\begin{equation}
  V_{\rm out} = G(T_{\rm in}, T_{\rm noise})\times\left(T_{\rm in}+T_{\rm noise}\right),
  \label{eq:vout}
\end{equation}
where $T_{\rm in}$ refers to any input signal, could be either $T_{\rm sky}$ or $T_{\rm ref}$, $V_{\rm out}$ is the corresponding voltage output, $T_{\rm noise}$ is the corresponding noise temperature and $G(T_{\rm in}, T_{\rm noise})$ is the photometric calibration constant (detector gain) which may depend on the input and noise temperatures (non-linear response) in general. The parametrization of the non-linear response model has been provided in~\cite{Mennella2009}
\begin{eqnarray}
  &&V_{\rm out} = G_{\rm tot}\left(T_{\rm in}+T_{\rm noise}\right)\,,\nonumber\\
  &&G_{\rm tot} =  \frac{G_0}{1+k G_0\left(T_{\rm in}+T_{\rm noise}\right)} \,,
  \label{eq:vout_compact}
\end{eqnarray}
$k$ is the photometric calibration constant, and have been estimated with various sky-load temperature for the 30 and 44 GHz receivers. If $k=0$ the response parameter (receiver gain), $G_{\rm tot}$, reduces to the constant $G_0$, and $k=\infty$ corresponds to infinite compression of the input signal. In this model, $G_0$ is proportional to gains of front-end module, band-end module, and the bandwidth.

The linearity of the receiver response is important for the in-flight calibration. For example, photometric calibration is performed by exploiting the dipole anisotropy and for the beam measurements through observations of bright point sources, e.g. Jupiter and Saturn.

The variation on $G$, $\delta G / G$, caused by a variation of the input temperature $\delta T$  can be calculated as

\begin{equation}
    \frac{\delta G}{G} = - \frac{k\, \delta T \, G_0}{1 + k\, G_0 ( T_{\rm sky} + \delta T + T_{\rm noise})}\,.
\end{equation}
For dipole anisotropy, $\delta T \sim \pm 3$~mK, and for bright point source like Jupiter, $\delta T\sim \pm 50$~mK. The estimation of the receiver parameters $G_0$, $T_{\rm noise}$ and $k$ can be found in~\cite{Mennella2009}. The non-linearity or the relative change on the detector gain has been estimated to be

\begin{equation}
\begin{array}{l l l}
    \frac{\delta G}{G} \lesssim 6\times 10^{-5} &\mbox{ for } &\delta T\sim \pm 3\, \mbox{mK} \\
    \mbox{}\\
    \frac{\delta G}{G} \lesssim 10^{-3} &\mbox{ for } &\delta T\sim\pm 50\, \mbox{mK},
\end{array}
\end{equation}
With the above limits on the gain, we can now calculate the non-linear response of temperature field $b$ of our parametrization by noting that $\delta G/G=bT/(1+a+bT)$. For $T\sim \pm 3$ mK this gives $b \lesssim 0.1$. We can see that during nominal operation with small input signal, the performance of the receiver is sufficiently linear as not to contaminate the bispectrum measurement. For bolometer detectors like HFI, the dynamic range is much smaller. This is usually not a problem if low temperatures are considered, but for the simultaneous measurement of galaxy and the CMB the non-linearity parameter $b>O(10)$ might pose a problem. To our knowledge, there is no available literature which characterizes the non-linearities of the HFI instrument~\cite{Rosset2010,Lamarre2010,Pajot2010,Ade2010,Maffei2010}. Since HFI will be the driving instrument for primordial bispectrum measurements, it is essential that instrumental non-linearities satisfy our benchmark criteria.

\section{Summary}
\label{sec:summary}

Future CMB experiments will reach the raw sensitivity to detect local shape $\fnl \sim 5 (1\sigma)$ using CMB temperature anisotropy information, and $\fnl \sim 2 (1\sigma)$ using combined the temperature and $E$-polarization information. However to reach this Fisher limit and reliably infer the level of primordial non-Gaussianity, precise and realistic calculations of the effect instrumental systematics on bispectrum are required. Although the impact of instrumental systematics have been extensively studied for the CMB power-spectrum, the effect on bispectrum have not been extensively explored. We have studied, in a fairly general manner, the scientific impact of the instrumental systematic effects on the primordial non-Gaussianity measurement from the CMB. For the first time, we introduce parametrization for non-linearities of the instrument for both temperature and polarization systematics. We have shown that although linear systematics fields can distort existing primordial bispectrum, they do not generate new spurious bispectrum. Perhaps more important, we show that certain types of second order non-linear systematics {\em generate} spurious bispectrum even if the primordial CMB is perfectly Gaussian.  Quantitatively, we propagate the effect of systematics errors to the CMB bispectrum and then to eventually assess their impacts on the non-Gaussianity parameter $\fnl$.

We consider different bispectrum configurations which are predicted by various well-motivated inflation models: local, equilateral, and orthogonal shape of non-Gaussianity. For each type of non-Gaussianity, we calculate the effect of instrumental systematics on the temperature bispectrum $TTT$, and bispectrum involving polarization fields: $EEE, TEE,$ and $TTE$. We discuss that by optimally designing the scan strategy, the level of several polarization systematics can be reduced. However, the quadrupole leakage can not be averaged out even with perfect coverage of rotation angles. Therefore, for polarization systematics we primarily focus on non-linear quadrupole leakage $\tilde q$. 

We calculate the tolerance limits on the systematics parameters so that spurious $f^{syst.}_{\rm NL}$ does not degrade our ability to constrain primordial non-Gaussianity. We discussed the implication of our findings to future CMB experiments in particular the space-based PLANCK mission. We find that the non-linear systematics does not significantly affects CMB power spectrum measurement. The effective local $f^{syst.}_{NL}$ could reach $O(10)$ before systematic effects show up in the CMB dipole measurement. As a result, if the non-linear systematics is not controlled by dedicated calibration, the measured bispectrum could be biased. The effect of linear systematics on the primordial bispectrum is small. For linear systematics control of {\it rms} $\sim10\%$, measurements at $\ell \lesssim 2000$ do not suffer significant degradation, however for experiments probing $\ell_{max} \gtrsim 2000$, even linear systematics could alter the cosmological bispectrum. Since secondary anisotropies start to dominate at around $\ell \sim 2000$, we conclude that linear systematics is much smaller worry than the nonlinear systematics.

\acknowledgments{We are particularly thankful to Matias Zaldarriaga for useful discussions on the project and providing comments on the paper. MS acknowledges helpful conversations with Xingang Chen. A.P.S.Y. gratefully acknowledges support from IBM Einstein fellowship and funding from NASA award number NNX08AG40G, NSF grant number AST-0807444. We acknowledge a CMBPol workshop held at University of Chicago in July 2009 where this work got started. 
\bibliography{myreferences}

\appendix

\section{Discussion on Polarization Systematics}
\label{sec:Stokes}

In the paper we mostly focused on the Jones parametrization formalism and generalized the formalism presented in Ref~\cite{HHZ}. However for the PSB it is convenient to parametrize systematics starting from Stokes parameters~\cite{Masi2006} $I$, $Q$ and $U$ 
($V=0$ for CMB polarization) as opposed to the Jones matrix formalism which 
is particularly useful for describing coherent polarimeters~\cite{HHZ,O'Dea:2006di}.  Systematics parameters in this formalism are: the gain factor $g$, the differential
beam-width of the beams $\mu$, the differential pointing $\rho$, 
the beam ellipticity $e$ and the beam rotation $\varepsilon$. These parameters are derived from differences in intensity in the 
Gaussian 2-D polarized beam response function for each polarization.
Although throughout this paper we employed the Jones formalism, it is constructive to compare it to the Stokes formalism ~\cite{SKPH} formalism.  This will shed light on the similarities and differences between the two. Table III summarizes the correspondence between the parameters of two formalism. 

Our starting point is the Fourier-space analysis of ~\cite{SKPH}. We expand the leading terms to first order and 
identify each one of them with the corresponding term of the Jones Formalism ~\cite{HHZ}. Since in real space the temperature and polarization patterns are convolved with the beams, these expressions are simply the product of their Fourier transforms in 
Fourier space. We restrict the discussion to an elliptical Gaussian beam (with major and minor axes $\sigma_{x}$ and $\sigma_{y}$)
\begin{eqnarray} 
B({\bf x})=\frac{1}{2\pi\sigma_{x}\sigma_{y}}\exp\left(-\frac{(x-\rho_{x})^{2}}{2\sigma_{x}^{2}}
-\frac{(y-\rho_{y})^{2}}{2\sigma_{y}^{2}}\right)
\label{eqn:unify_beam}
\end{eqnarray}
and its Fourier transform is
\begin{eqnarray}
\tilde{B}({\bf l})=
\exp\left(-\frac{l_{x}^{2}\sigma_{x}^{2}}{2}-\frac{l_{y}^{2}\sigma_{y}^{2}}{2}
+i{\bf l}\cdot{\bf\rho}\right).
\label{eqn:unify_beam_fourier}
\end{eqnarray}
The pointing error merely shifts the phase of the beam representation 
in Fourier-space.

From Eq.~(12) of ~\cite{SKPH} the beam function with ellipticity $e$ and pointing $\rho$ in $l$-space reads
\begin{eqnarray}
\tilde{B}({\bf l})&=&e^{-y}\sum_{n=-\infty}^{\infty}\sum_{m=-\infty}^{\infty}i^{2m+n}I_{m}(z)J_{n}(l\rho) \times e^{i(2m+n)\psi-in\theta}e^{i(2m+n)(\phi_{l}-\alpha)}
\end{eqnarray}
where
\begin{eqnarray}
y&\equiv &\frac{l^{2}}{4}(\sigma_{x}^{2}+\sigma_{y}^{2})\nonumber\\
z&\equiv &\frac{l^{2}}{4}(\sigma_{x}^{2}-\sigma_{y}^{2})
\end{eqnarray}
and $e=\frac{\sigma_{x}-\sigma_{y}}{\sigma_{x}+\sigma_{y}}$ and the pointing is defined by $\rho_{x}=\rho\cos\theta$ 
and $\rho_{y}=\rho\sin\theta$. The angle $\alpha$ is scanning strategy- and time-dependent and is shown in Figure 6.

\begin{figure*}[t]
\includegraphics[width=120mm,angle=0]{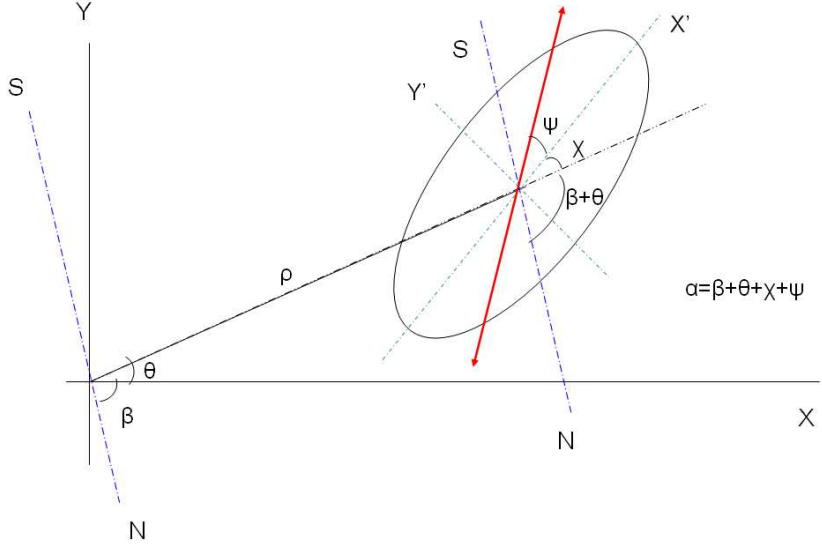}
\caption{Angles layout: The X-Y system is fixed to the focal plane, which is itself rotating with an angle $\beta$ 
with respect to the fixed N-S system on the sky. The X'-Y' frame coincides with the ellipse principal axes and the angle 
$\psi$ represents the angle that the polarization-sensitive-axis makes with the ellipse major axes. 
In the presence of a half wave plate (HWP) the angle $\psi$ varies, otherwise it should be fixed. The pointing is 
represented by the 2D vector ${\bf \rho}$ which makes an angle $\theta$ with the positive X-axis. The angle $\chi$ 
represents the tilt of the ellipse major axis with respect to ${\bf \rho}$.}
\end{figure*}

The observed CMB polarization obtained as in Eq.~(18) of ~\cite{SKPH}
\begin{eqnarray}
Q'\pm iU'&=&\frac{1}{D}\langle B_{+}\rangle(Q\pm iU)
+\frac{2}{D}\langle B_{-}e^{\pm 2i\alpha}\rangle T
+\frac{1}{D}\langle B_{+}e^{\pm 4i\alpha}\rangle(Q\mp iU)\nonumber\\ 
&-&\frac{1}{D}\langle B_{+}\rangle\langle e^{\pm 4i\alpha}\rangle(Q\mp iU)
-\frac{2}{D}\langle B_{-}e^{\pm 2i\alpha}\rangle \langle e^{\pm 4i\alpha}\rangle T
-\frac{1}{D}\langle B_{+}e^{\mp 4i\alpha}\rangle\langle e^{\pm 4i\alpha}\rangle(Q\pm iU)
\label{eqn:unifystokes}
\end{eqnarray}
where angular brackets $\langle ...\rangle$ represent averaging over `hits' at the given sky-pixel. 
A scanning strategy is said `ideal' if $\langle e^{\pm in\alpha}\rangle=0$ for any $n\neq 0$ and is 
`uniform' if $\langle e^{\pm in\alpha}\rangle$ does not vanish, yet it is constant across the sky.
The symbols $B_{\pm}$ stand for $\frac{1}{2}(B_{1}\pm B_{2})$ where $B_{1}$ and $B_{2}$ are the first and 
second beams in the beam pair.
The map inversion becomes singular when $D\equiv 1-\langle e^{4i\alpha}\rangle \langle e^{-4i\alpha}\rangle$ 
vanishes. This implies that a good scanning strategy should minimize $\langle e^{4i\alpha}\rangle$ and 
$\langle e^{-4i\alpha}\rangle$, terms responsible spin-flip, i.e. $Q\pm iU\rightarrow Q\pm iU$, since 
$\langle e^{4i\alpha}\rangle$ are spin-$\pm 4$ fields. We therefore neglect the second line of Eq.(\ref{eqn:unifystokes}), 
noting that these terms need not vanish in general and approximate $D\approx 1$ to first order in scanning 
strategy quantities. Note also that we ignore pixel-rotation for the 
sake of simplicity since at the leading order it decouples from other beam systematics and can be 
accounted for by simply setting
\begin{eqnarray}
Q'\pm iU'=e^{\pm 2i\varepsilon}(Q\pm iU).
\end{eqnarray}
The analog of Eq.(\ref{eqn:unifystokes}) in the ~\cite{HHZ} formalism reads (Eqs. 17 \& 18 of ~\cite{HHZ})
\begin{eqnarray}
Q'\pm iU'&=&Q\pm iU+\sigma ({\bf p}\cdot{\bf \nabla})(Q\pm iU)+\sigma(d_{1}+id_{2})(\partial_{1}\pm i\partial_{2})T\nonumber\\
&+&\sigma^{2}q(\partial_{1}\pm i\partial_{2})^{2}T+(a+2i\omega)(Q\pm iU)+(f_{1}+if_{2})(Q\pm iU)+(\gamma_{1}\pm i\gamma_{2})T.
\end{eqnarray} 
Note, in particular, that since $T$ is a spin-$0$ field it must couple to spin$\pm 2$ to generate $Q'\pm iU'$. Now, since 
$\partial_{1}\pm i\partial_{2}$ is a spin$\pm 1$, $d_{1}+id_{2}$ must be a spin$\pm 1$ field which can come from a 
non-vanishing dipole-moment of the scanning strategy $\langle e^{\pm i\alpha}\rangle$. Similarly, $\partial_{1}\mp i\partial_{2}$ is a spin$\mp 1$ field and therefore must couple to a spin$\pm 3$ field, such as the octupole moment of the scanning strategy, to form the required spin$\pm 2$ field for temperature leakage to polarization. Expanding Eq.(\ref{eqn:unifystokes}) at leading order we can identify the ~\cite{HHZ} with the corresponding ~\cite{SKPH} terms. This is summarized in Table III.

Note also that the scanning strategy functions, e.g. $\langle\cos 4\alpha\rangle$, can be space-dependent and very much like to CMB lensing this may induce higher-order correlations. However, since the scanning strategy is completely uncorrelated with the underlying temperature and polarization this will show up only at the trispectra level and is therefore irrelevant for the current work.
As mentioned above, the main difference between the formalisms stems from the different treatment of the polarization direction $\psi+\chi$ with respect to the pointing (Fig. 6). A good example where this difference can show up is the inclusion of a HWP in the experimental setup. Averaging over the angle $\psi$ by employing a fast rotating HWP will result in no polarization in the ~\cite{SKPH} formalism while the parametrization adopted by ~\cite{HHZ} fixes $\psi+\chi$ to 0, i.e. the polarization sensitive directions are aligned with one of the ellipse principal axes. As has been shown by ~\cite{SKPH} the level of systematic $B$-mode (for a given beam ellipticity) depends on $\psi+\chi$ and it maximizes for $\psi+\chi=45^{\circ}$. In general, these two formalisms agree upon making the appropriate identifications of the model parameters.

\begin{table}[t]
\begin{tabular}{l l||l l}
\hline
\bf{HHZ}& &\bf{SKPH}& \\
\hline
Calibration & $a$ & Gain &$g$\\
Rotation & $\rot$ & Pixel rotation & $\varepsilon$\\
Pointing &$p_a$& Pointing & $\frac{\rho_x}{2\sigma}\langle \cos(\alpha)\rangle
+\frac{\rho_y}{2\sigma}\langle \sin(\alpha)\rangle$\\
Pointing &$p_b$& Pointing & $\frac{\rho_x}{2\sigma}\langle \sin(\alpha)\rangle
-\frac{\rho_y}{2\sigma}\langle \cos(\alpha)\rangle$\\
Flip &$f_a$& Flip &$\langle\cos 4\alpha\rangle$\\
Flip &$f_b$& Flip &$\langle\sin 4\alpha\rangle$\\
Monopole &$\gamma_a$ & Diff. gain &$g\langle \cos 2\alpha\rangle$\\
Monopole &$\gamma_b$ & Diff. gain &$g\langle \sin 2\alpha\rangle$\\
Dipole &$d_a$& Diff. pointing &$\frac{\rho_{x}}{\sigma}\langle e^{i\alpha}\rangle$\\
Dipole &$d_b$& Diff. pointing &$-\frac{\rho_{y}}{\sigma}\langle e^{i\alpha}\rangle$\\
Quadrupole &$q$& Diff. ellipticity &$e\cdot e^{2i\psi}$\\
\hline
\end{tabular}
\caption{Comparison between the ~\cite{HHZ} and  ~\cite{SKPH} formalisms: the various parameters used in the two formalisms are identified. 
It is shown that most of beam systematics vanish in case of ideal scanning strategy (pointing, spin-flip, monopole, dipole). 
The respective names and parameters are those used in the ~\cite{HHZ} and ~\cite{SKPH} papers.}
\end{table}

Although we have considered a pretty general class of instrumental systematics, there are several effects one may want to include to extend our analysis: leakage of galactic temperature signal to the polarization maps due to bandpass mismatch, beam side-lobes and variations in instrumental response with frequency. In our instrumental systematics model, we have assumed monochromatic response of the detector. Another concern may be that our toy systematic model does not capture all the realistic effects. For example, pointing errors would generally introduce polarization leakage regardless of the shape of the beam. A good model for the pointing error depends on the scanning strategy, as the pointing offset could be, in principle, strongly correlated for which the error keeps constant over a period of time. The pointing error would cause the misplacement of time-ordered-data samples into wrong pixels during map-making. In other words, the error increases the effective size of the pixels with extra smoothing effect (related to the rms of the pointing error), making every pixel on the produced CMB map have its own effective beam. 

If the bias of bispectrum introduced by instrumental systematics is noticeable, it is possible to correct for the instrumental systematic effects by estimating the bias induced by the systematics if we have a precise enough knowledge of them. One way to do this is to use the recovered temperature and $E$-mode maps as input sky and simulate the instrument using the known beams. The output of this simulation contains some non-Gaussianity, e.g. certain type of bispectrum, not present initially, which is an estimate of the observed spurious bispectrum. The efficiency of this correction obviously depends on how well the systematics are known and modeled. 

\section{Full-sky Treatment}

In this section, we focus on the derivation of the systematics distorted/generated CMB bispectrum under the more appropriate spherical sky treatment. The full-sky formalism can be obtained by simply replacing the Fourier components with spherical harmonic multipole moments. We follow the mathematical notation and structure presented in~\cite{Hu2000,Cooray2008}
\begin{eqnarray}
\cmb^{obs}_{l m} &\approx& \cmb_{l m} + \int d\bn Y_l^{m*} a(\bn) \cmb(\bn) + \int d\bn Y_l^{m*} b(\bn) \cmb(\bn)\cmb(\bn) \nonumber\\
& = & \cmb_{l m} + \sum_{\lp \mpr} \sum_{\ldp \mdp}  \cmb_{\ldp \mdp}
			\bigg[ a_{\lp \mpr}  I_{l \lp \ldp}^{m \mpr \mdp} + \sum_{\ltp \mtp} b_{\ltp \mtp} J_{l \lp \ldp \ltp}^{m \mpr \mdp \mtp}  \cmb_{\ltp \mtp} \bigg] \,, 
\end{eqnarray}
note that the integrals over the spherical harmonics are replaced by the geometrical factors defined in the following
\begin{eqnarray}
I_{l \lp \ldp}^{m \mpr \mdp} &=&
			\int d\bn \, Y_l^{m*}  Y_{\lp}^{\mpr} Y_{\ldp}^{\mdp} \,, \\
\label{eqn:IJform}
J_{l \lp \ldp \ltp}^{m \mpr \mdp \mtp}
			 &=& \int d\bn\, Y_l^{m*}  Y_{\lp}^{\mpr} Y_{\ldp}^{\mdp} Y_{\ltp}^{\mtp}      \,.
\end{eqnarray}

The CMB temperature bispectrum with systematics contaminations can then be expressed as
\begin{eqnarray}
B^{obs}_{l_1 l_2 l_3} &=& \sum_{m_1 m_2 m_3} \wthrj{l_1}{l_2}{l_3}{m_1}{m_2}{m_3}  \langle \cmb^{obs}_{l_1 m_1} \cmb^{obs}_{l_2 m_2} \cmb^{obs}_{l_3 m_3} \rangle  \, ,
\end{eqnarray}
where the average is over the CMB and systematics field realizations. leading to
\begin{eqnarray}
&& {B}_{l_1 l_2 l_3}^{obs} = \sum_{m_1 m_2 m_3}  \wthrj{l_1}{l_2}{l_3}{m_1}{m_2}{m_3}  \bigg[  \langle {\cmb}_{l_1 m_1} {\cmb}_{l_2 m_2} {\cmb}_{l_3 m_3} \rangle  + \sum_{\lthrp \mthrpr} \sum_{\lthrdp \mthrdp} \sum_{\lthrtp \mthrtp} \langle{\cmb}_{l_1 m_1} {\cmb}_{l_2 m_2}  {\cmb}_{l_3 m_3}\cmb_{\lthrdp \mthrdp} b_{\lthrtp \mthrtp} \rangle J_{l_3 \lthrp \lthrdp \lthrtp}^{m_3 \mthrpr \mthrdp \mthrtp}  +{\rm 2 \; Perm.}\nn 
&& + \sum_{\ltwop \mtwopr} \sum_{\ltwodp \mtwodp}  \sum_{\lthrp \mthrpr} \sum_{\lthrdp \mthrdp} \langle {\cmb}_{l_1 m_1}  \cmb_{\ltwodp \mtwodp} \cmb_{\lthrdp \mthrdp} \hspace{3mm} a_{\ltwop \mtwopr} a_{\lthrp \mthrpr} \rangle I_{l_2 \ltwop \ltwodp}^{m_2 \mtwopr \mtwodp} I_{l_3 \lthrp \lthrdp}^{m_3 \mthrpr \mthrdp} + {\rm 2 \; Perm.} \bigg] \, . 
\end{eqnarray}
Here the first term is the desired primordial bispectrum, second terms is the bispectrum due to non-linear response of the instrument and the last term is the distorted bispectrum due to linear systematics $a$. We should the details for the linear-systematics, but can similarly be applied to the non-linear response. Noting that the Wigner-3$j$ symbol obeys the identity
\begin{equation}
\sum_{m_1 m_2} \wthrj{l_1}{l_2}{l_3}{m_1}{m_2}{m_3}  \wthrj{l_1}{l_2}{\lthrdp}{m_1}{m_2}{\mthrdp} = \frac{ \delta_{l_3 \lthrdp}\delta_{m_3 \mthrdp}}{(2l_3+1)},
\end{equation}
we can re-write the distorted bispectrum as 
\begin{eqnarray}
\tilde{B}_{l_1 l_2 l_3}^{\cmb} &=& {B}_{l_1 l_2 l_3}^{\cmb} +  \sum_{\ltwodp \lthrdp} B_{l_1  \ltwodp \lthrdp}^{\cmb} \sum_{\ltwop} C_{\ltwop}^{aa} R  + 2 \; \text{Perm.}
\label{eq:lensedbis}
\end{eqnarray}
where
\begin{eqnarray}
R =  \sum_{\mtwopr \mtwodp \mthrdp}  \sum_{m_1 m_2 m_3} \wthrj{l_1}{l_2}{l_3}{m_1}{m_2}{m_3} \wthrj{l_1}{\ltwodp}{\lthrdp}{m_1}{\mtwodp}{\mthrdp}\times I_{l_2 \ltwop \ltwodp}^{m_2 \mtwopr \mtwodp} I_{l_3 \ltwop \lthrdp}^{m_3 \mtwopr \mthrdp} 
\end{eqnarray}
In order to evaluate $R$, we first re-express $I_{l \lp \ldp}^{m \mpr \mdp}$ as
\begin{equation}
I_{l \lp \ldp}^{m \mpr \mdp} = f_{l \lp \ldp} \wthrj{l}{\lp}{\ldp}{m}{\mpr}{\mdp} \, ,
\end{equation}  
where
\begin{equation}
f_{l \lp \ldp} = \wthrj{l}{\lp}{\ldp}{0}{0}{0} \,.
\end{equation} 
Then the expression for $R$ can be re-written as
\begin{eqnarray}
R &=& f_{l_2 \ltwop \ltwodp} \; f_{l_3 \ltwop \lthrdp} \sum_{\mtwopr \mtwodp \mthrdp}  \sum_{m_1 m_2 m_3} \wthrj{l_1}{l_2}{l_3}{m_1}{m_2}{m_3}\wthrj{l_1}{\ltwodp}{\lthrdp}{m_1}{\mtwodp}{\mthrdp} \wthrj{l_2}{\ltwop}{\ltwodp}{m_2}{\mtwopr}{\mtwodp} \wthrj{l_3}{\ltwop}{\lthrdp}{m_3}{\mtwopr}{\mthrdp}  f_{l_2 \ltwop \ltwodp} \; f_{l_3 \ltwop \lthrdp} (-1)^{\lp+\ltwop+\lthrp} \wsixj{l_1}{l_2}{l_3}{\ltwop}{\lthrdp}{\ltwodp} \, , \nonumber \\
\end{eqnarray} 
where,  in  the  last  step,  we  have
introduced  the Wigner-6$j$  symbol.  The  values of  the
Wigner-6$j$  symbol  can  be  computed  numerically with  a  fast  and
efficient recursive algorithm.

Finally,  substituting   the  expressions  for  $R$ and including all permutations in a single expression, we can write the systematics distorted bispectrum as
\begin{eqnarray}
&& \tilde{B}_{l_1 l_2 l_3}^{\cmb} = B_{l_1 l_2 l_3}^{\cmb} + \sum_{lpq} C_{\ell}^{SS}\bigg[ f_{l_2 l p}f_{l_3 l q} (-1)^{n} \wsixj{l_1}{l_2}{l_3}{l}{q}{p} {B}_{l_1 p q}^{\cmb} f_{l_3 l p}f_{l_1 l q} (-1)^{n} \wsixj{l_1}{l_2}{l_3}{p}{l}{q} {B}_{p l_2 q}^{\cmb} f_{l_1 l p}f_{l_2 l q} (-1)^{n} \wsixj{l_1}{l_2}{l_3}{q}{p}{l} {B}_{p q l_3}^{\cmb}  \bigg] \nonumber \\
\end{eqnarray}
where  $n\equiv (l+p+q)$.

\end{document}